\documentclass[prx,twocolumn,floatfix,superscriptaddress,amsmath,aps,longbibliography]{revtex4-2}
\pdfoutput=1
\usepackage[T1]{fontenc}\usepackage[latin1]{inputenc}
\usepackage{dcolumn,graphicx,color,booktabs,microtype,afterpage} 
\usepackage[charter,greekuppercase=italicized]{mathdesign}\usepackage{sidecap}
\usepackage{amsmath}
\usepackage[colorlinks,plainpages=false,linkcolor=blue,urlcolor=blue,citecolor=blue,pdfpagemode=UseNone,pdfstartview=FitBH]{hyperref}
\graphicspath{{Figures/}}
\newcommand{\dso}{DyScO$_3$}
\newcommand{\cco}{Ca$_3$Co$_2$O$_6$}
\newcommand{\dto}{Dy$_2$Ti$_2$O$_7$}
\newcommand{\TN}{$T_{\mathrm{N}}$}

\begin{document}
%
%

\title{Slow spin dynamics and quantum tunneling of magnetization in the dipolar antiferromagnet \dso}
\author{N.~D.~Andriushin}
\affiliation{Institut f{\"u}r Festk{\"o}rper- und Materialphysik, Technische Universit{\"a}t Dresden, D-01069 Dresden, Germany}
\author{S.~E.~Nikitin}
\affiliation{Paul Scherrer Institute (PSI), CH-5232 Villigen, Switzerland}
\author{G.~Ehlers}
\affiliation{Neutron Technologies Division, Oak Ridge National Laboratory, Oak Ridge, Tennessee 37831, USA}
\author{A.~Podlesnyak}
\affiliation{Neutron Scattering Division, Oak Ridge National Laboratory, Oak Ridge, Tennessee 37831, USA}

\begin{abstract}
We present a comprehensive study of static and dynamic magnetic properties in the Ising-like dipolar antiferromagnet (AFM) \dso\ by means of DC and AC magnetization measurements supported by classical Monte-Carlo calculations. Our AC-susceptibility data show that the magnetic dynamics exhibit a clear crossover from an Arrhenius-like regime to quantum tunneling of magnetization (QTM) at $T^* = 10$~K. Below $T_{\mathrm{N}} = 3.2$~K \dso\ orders in an antiferromagnetic $GxAy$-type magnetic structure and the magnetization dynamics slow down to the minute timescale. The low-temperature magnetization curves exhibit complex hysteretic behavior, which depends strongly on the magnetic field sweep rate. We demonstrate that the low-field anomalies on the magnetization curve are related to the metamagnetic transition, while the hysteresis at higher fields is induced by a strong magnetocaloric effect. Our theoretical calculations, which take into account dipolar interaction between Dy$^{3+}$ moments, reproduce essential features of the magnetic behavior of \dso. We demonstrate that \dso\ represents a rare example of inorganic compound, which exhibits QTM at a single-ion level and magnetic order due to classical dipolar interaction. 
\end{abstract}

\maketitle

\section{Introduction}
\begin{figure}
  \includegraphics[width=0.8\linewidth]{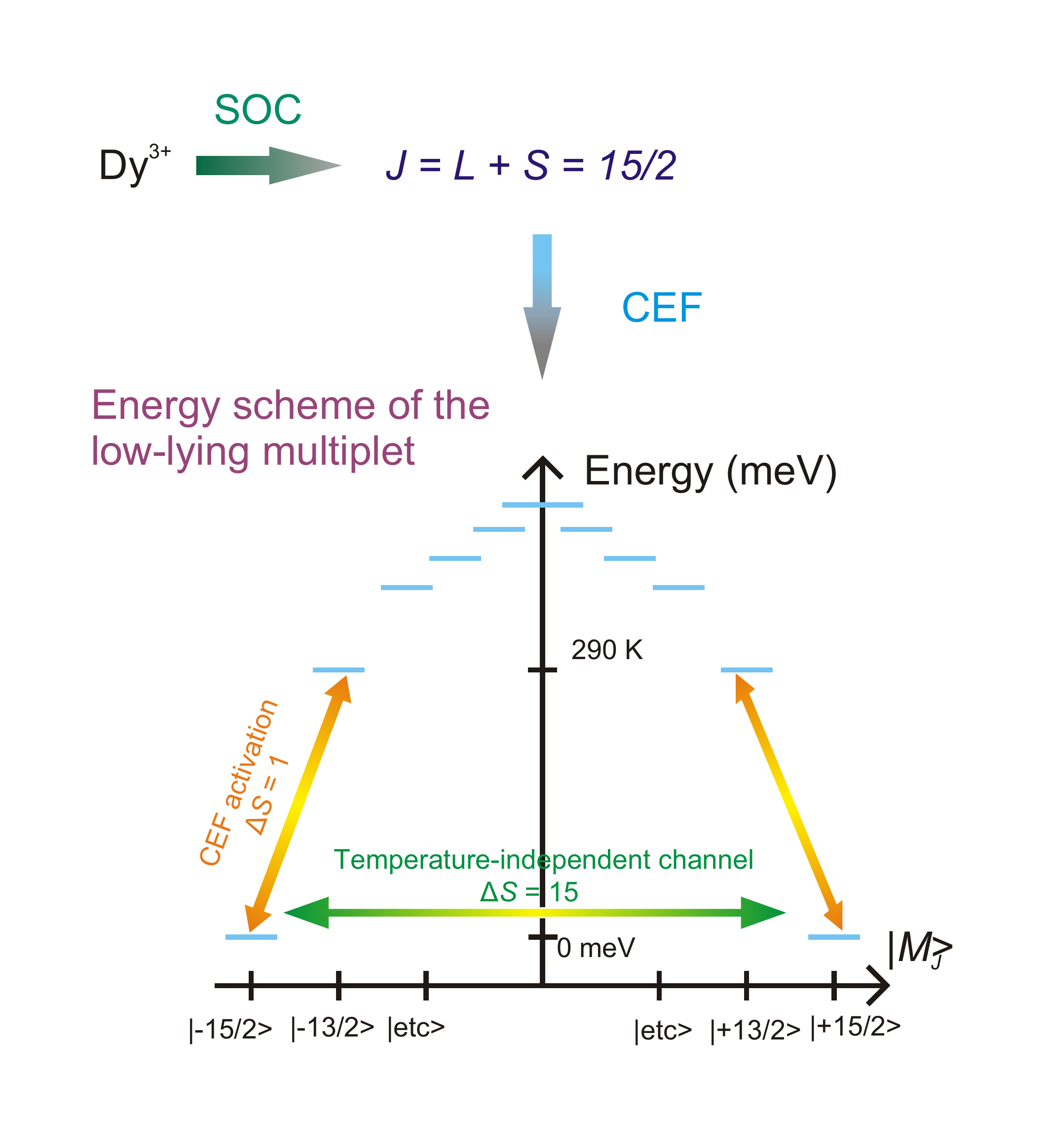}
  \caption{~Sketch of the energy diagram of $\mathbf{J} = 15/2$ multiplet of Dy$^{3+}$ in \dso, shown in $|M_J\rangle - E$ coordinates. Spin-orbit coupling (SOC) produces the $\mathbf{J} = 15/2$ multiplet, which is split into 8 doublets by CEF. The ground state doublet consists of $|\pm15/2\rangle$ wavefunctions, following by excited $|\pm13/2\rangle$, $|\pm11/2\rangle$, etc.  While the temperature dependence of the spin excitation associated with $|\pm15/2\rangle \rightarrow |\pm13/2\rangle$ transition follows Arrhenius law, the direct transition between $|\pm15/2\rangle$ states is the temperature-independent QTM process.}
  \label{CEF_scheme}
\end{figure}

Timescale of spin dynamics -- the time required to flip a single spin -- in conventional magnetic materials is of the order of femto- to picosecond~\cite{de2011laser,neeraj2021inertial, de2021colloquium}. Anomalous slowing down by multiple orders of magnitude down to the millisecond range is known take place in some single-molecule magnets and is induced by strong uniaxial anisotropy~\cite{gatteschi2003quantum,langley2014modulation, guo2011relaxation}. In that case, the strong crystalline electric field (CEF) splits the ground-state multiplet $\mathbf{J}$ of a magnetic ion and creates a doublet ground state, which consists of two states with maximal projection of angular momentum, pointing in the opposite directions, $|\psi^0_{\pm}\rangle = |\pm{} J\rangle$. Thus, the direct transition between  $|\psi^0_{+}\rangle$ and  $|\psi^0_{-}\rangle$ states requires a change of the total momentum $\Delta{}J$ by more than one, and thus is forbidden by selection rules of many conventional single-(quasi)particle emission/absorption processes. Therefore, the matrix element for this transition is low, see Fig.~\ref{CEF_scheme}.

In this case there are two possible ways to change the spin momentum: (i) via an activation process to the excited doublet with wavefunctions $|\psi^1_{\pm}\rangle = | \pm{} J \mp 1 \rangle$; or (ii) as a direct transition between $|\psi^0_{+}\rangle$ and $|\psi^0_{-}\rangle$ via the quantum tunneling of magnetization (QTM) process~\cite{gatteschi2003quantum}. The first mechanism dominates in the temperature range $T \gtrsim \Delta_{\mathrm{CEF}}/k_{\mathrm{B}}$ ($\Delta_{\mathrm{CEF}}$ is the energy gap to the excited doublet and $k_{\mathrm{B}}$ is the Boltzmann constant), but it becomes ineffective at lower temperature where the QTM dominates the magnetic relaxation.  

QTM is a well-known process in single-molecule magnets, however to the best of our knowledge, among inorganic crystals there are only few well-documented examples including \dto~\cite{matsuhira2001novel, snyder2004low, matsuhira2011spin, takatsu2013ac, gardner2011slow, ehlers2002dynamical} and \cco~\cite{hardy2004magnetic, hegde2020magnetic}. In both cases, the strong CEF produces large uniaxial anisotropy, which freezes magnetic moments below a crossover temperature $T^* \approx 13$ and $T^* \approx\ 9$~K, for \dto\ and \cco\ respectively. However, in addition to the QTM both materials also show complex collective magnetic behavior, classical spin-ice physics in case of \dto\ and a frustrated spin-chain behavior in \cco, which obscure the QTM physics. 
Another prominent example is a dipolar ferromagnet LiHoF$_4$ and its diluted modifications LiHo$_x$Y$_{1-x}$F$_4$~\cite{bitko1996quantum,giraud2001nuclear, bertaina2006cross, johnson2012evolution}, where electronic and nuclear spins of Ho$^{3+}$ ions are coupled because of large hyperfine interaction, which produces complex slow dynamics at low temperatures.

In this work we focus on a classical Ising-like dipolar AFM \dso, whose magnetic properties were studied in details previously~\cite{wu2017magnetic, ke2009low, wu2019large, bluschke2017transfer}. It exhibits non-collinear AFM ordering below $T_{\mathrm{N}} = 3.2$~K with the propagation wavevector $\mathbf{k} = (001)$. Magnetization and neutron diffraction measurements show that Dy exhibits strong uniaxial anisotropy at low temperature and the easy-axis lies in the $ab$-plane, with a $\pm28^{\circ}$ angle to the [010] direction. Inelastic neutron scattering (INS) measurements show that the ground state doublet is well-isolated from the first excited doublet located at 290~K. Point-charge model calculations supported by magnetization measurements show that the wavefunction of the ground state doublet consists of almost pure $|\pm 15/2\rangle$ states, making \dso\ a prospective material to search for the QTM effect. 

In this work we performed a comprehensive study of the low-temperature magnetic behavior in \dso\ by means of AC and DC magnetization measurements. We observed a clear peak in imaginary part of dynamical spin susceptibility $\chi''$. It exhibits a crossover between an Arrhenius-like regime at high temperatures and a temperature-independent regime below $\lesssim\ 10$~K, which is a fingerprint of QTM behavior. The $M(B)$ curves taken below \TN\ demonstrate complex hysteretic behavior. By using classical Monte-Carlo simulations with dipolar interactions we reproduced essential features of magnetic behavior of \dso: (i) the type of magnetic ordering and the ordering temperature \TN; (ii) temperature dependences of the magnetic specific heat and magnetization; (iii) behavior of magnetic correlation length above the \TN; (iv) kink and a broad magnetic hysteresis on the $M(B)$ curves. Our results demonstrate that the low-temperature behavior of \dso\ is described by a combination of CEF-induced QTM, dipolar intersite interaction and the strong magnetocaloric effect.

\section{Results and Analysis}
\subsection{Slow dynamics and magnetic order at zero field}\label{app:SDMO}
\begin{figure}[tb!]
  \includegraphics[width=1.0\linewidth]{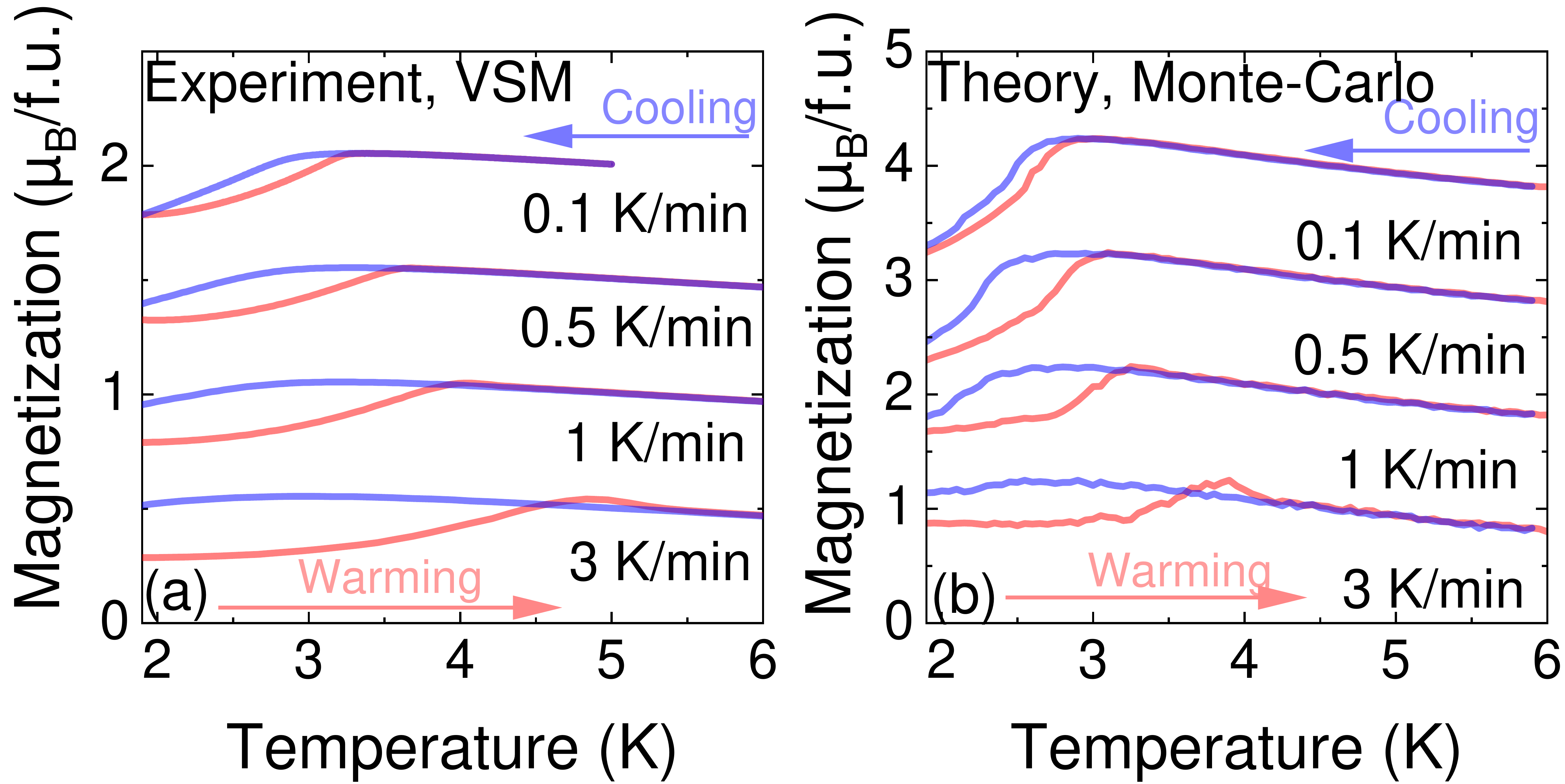}
  \caption{~Temperature dependences of the magnetization of \dso\ measured using VSM at $B = 0.1$~T (a) and calculated by Monte-Carlo (b) with different temperature sweep rates as described in Sec.~\ref{SubSec:MCS}. Red and blue lines represent data collected on warming and cooling respectively. Data are in panels (a) and (b) are shifted respectively by +0.5~$\mu_{\mathrm{B}}$/f.u. and +1~$\mu_{\mathrm{B}}$/f.u. vertically for clarity.}
  \label{M_vs_T_rate}
\end{figure}

We start the presentation of our results with the magnetization data collected as a function of temperature with different sweep rates as shown in Fig.~\ref{M_vs_T_rate}~(a). Note that the magnetic field was applied along the easy direction, $B \parallel\ [010]$ in all measurements and calculations. All curves collected upon warming up show a clear cusp anomaly associated with AFM ordering. Noticeably, the position of the cusp shifts with the sweep rate $dT/dt$. One can also see that the field-cooling (FC) curves differ considerably from those collected upon warming up: (i) the cusp associated with the AFM transition becomes less well-defined and is almost gone for $dT/dt \geq 1$~K/min; (ii) the FC and warming up curves show considerable hysteresis below \TN.

These results indicate the presence of a considerable magnetization relaxation at low temperatures, which takes place on a timescale of minutes. It is unexpected for a  conventional antiferromagnet, but rather reminiscent of the spin-glass (SG) state. At sufficiently low temperature, SG materials do not exhibit long-range magnetic order, but instead form magnetic clusters with short-range magnetic correlations~\cite{kroder2019spin, anand2012ferromagnetic}. A clear fingerprint of the SG behavior is a broad peak in the imaginary part of the AC-susceptibility, $\chi''(f)$, which should follow an Arrhenius-like temperature dependence over a broad temperature range~\cite{quilliam2008evidence}.

\begin{figure}[tb]
  \includegraphics[width=1.0\linewidth]{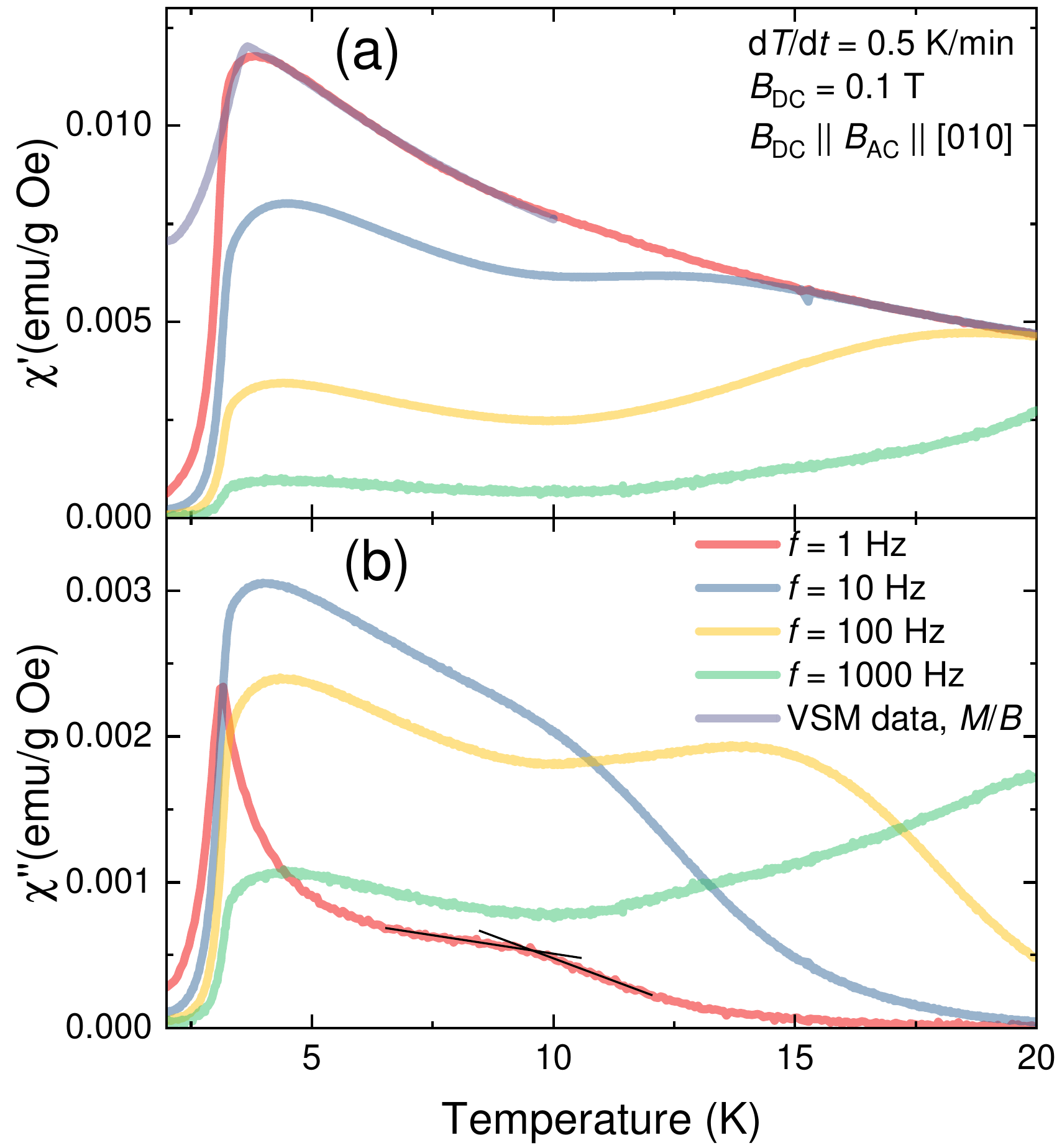}
  \caption{~Temperature dependence of real (a) and imaginary (b) parts of the complex longitudinal AC susceptibility of \dso\ measured at $B = 0.1$~T applied along [010] at different frequencies. The static spin susceptibility $M/B$ measured using VSM is shown in panel (a). Crossed black lines in panel (b) illustrate how the crossover temperature was determined. }
  \label{chi_vs_T}
\end{figure}

\begin{figure}[tb]
  \includegraphics[width=1.0\linewidth]{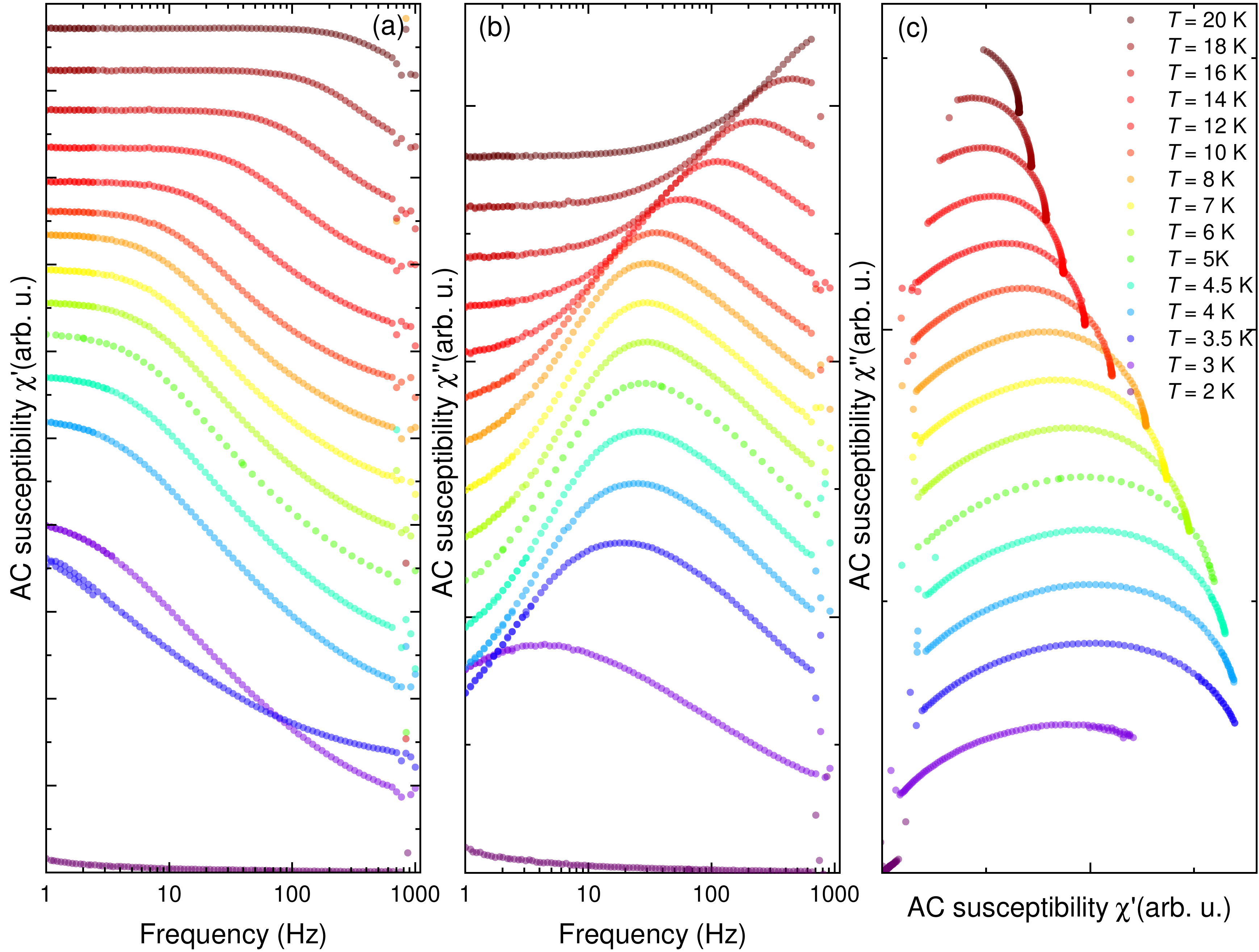}
  \caption{~Frequency dependence of the AC-susceptibility measured at multiple temperatures between 2 and 20~K. Panels (a) and (b) show the real [$\chi'(f)$] and imaginary [$\chi''(f)$] parts of the AC-susceptibility, respectively. Panel (c) shows the Cole-Cole plot $\chi''(\chi')$. All data are shown with a constant vertical offset for clarity. }
  \label{chi_vs_chi}
\end{figure}

To discuss \dso\ in this context, Figures ~\ref{chi_vs_T} and ~\ref{chi_vs_chi} display our temperature- and frequency-dependent AC-susceptibility measurements.
Figure~\ref{chi_vs_T}~(a) shows the temperature dependence of the real part of the AC-susceptibility, $\chi'(T)$, measured at different frequencies along with the static spin susceptibility, $M/B$, measured with VSM.  The curves collected at $f = 1$~Hz and at static regime agree well above \TN, and display a single peak at the transition temperature. The qualitative behavior changes when the frequency is increased. 
The low-temperature susceptibility measured at $f \geq\ 10$~Hz is reduced, but returns to $M/B$ above a frequency-dependent crossover temperature. The high-temperature tails of all curves follow the same Curie-Weiss law.

The  $\chi''(T)$ curve measured at 1~Hz shows a strong divergence at \TN, as expected for an AFM system, and a weak shoulder-like feature at $\sim 7$~K. With increasing frequency the shape of the peak at the ordering temperature changes significantly and becomes similar to the one observed in $\chi'(T)$. The second anomaly also shifts with frequency.  We quantified the positions of the second anomaly using the inflection point as shown for 1~Hz curve in Fig.~\ref{chi_vs_T}~(b). 

To further reveal the frequency dependence of the spin susceptibility in \dso\ we performed measurements of $\chi'(f)$ and $\chi''(f)$ at multiple temperatures, and the results are summarized in Fig.~\ref{chi_vs_chi}.  The $\chi'(f)$ demonstrates a plateau at low frequencies and a gradual decrease above temperature-dependent crossover frequency. The $\chi''(f)$ curves exhibit a strong broad peak at $T \geq 3$~K. The position of the peak shifts down with decreasing temperature, however between $T = 10$ and 4~K $\chi''(f)$ remains almost unchanged. When cooling below \TN\ the peak height decreases and shifts towards lower frequencies, which could not be followed further with our AC setup. Figure~\ref{chi_vs_chi}(c) shows Cole-Cole plots $\chi''(\chi')$ at different temperatures. For a system with a single relaxation channel (or symmetrical distribution of the relaxation channels), the curves should follow a semi-circular trajectory. However, the curves measured with \dso\ are asymmetric, indicating a more complex distribution of the relaxation times~\cite{havriliak1967complex,topping2018ac, miskinis2009havriliak}. 

\begin{figure}[tb!]
  \includegraphics[width=1.0\linewidth]{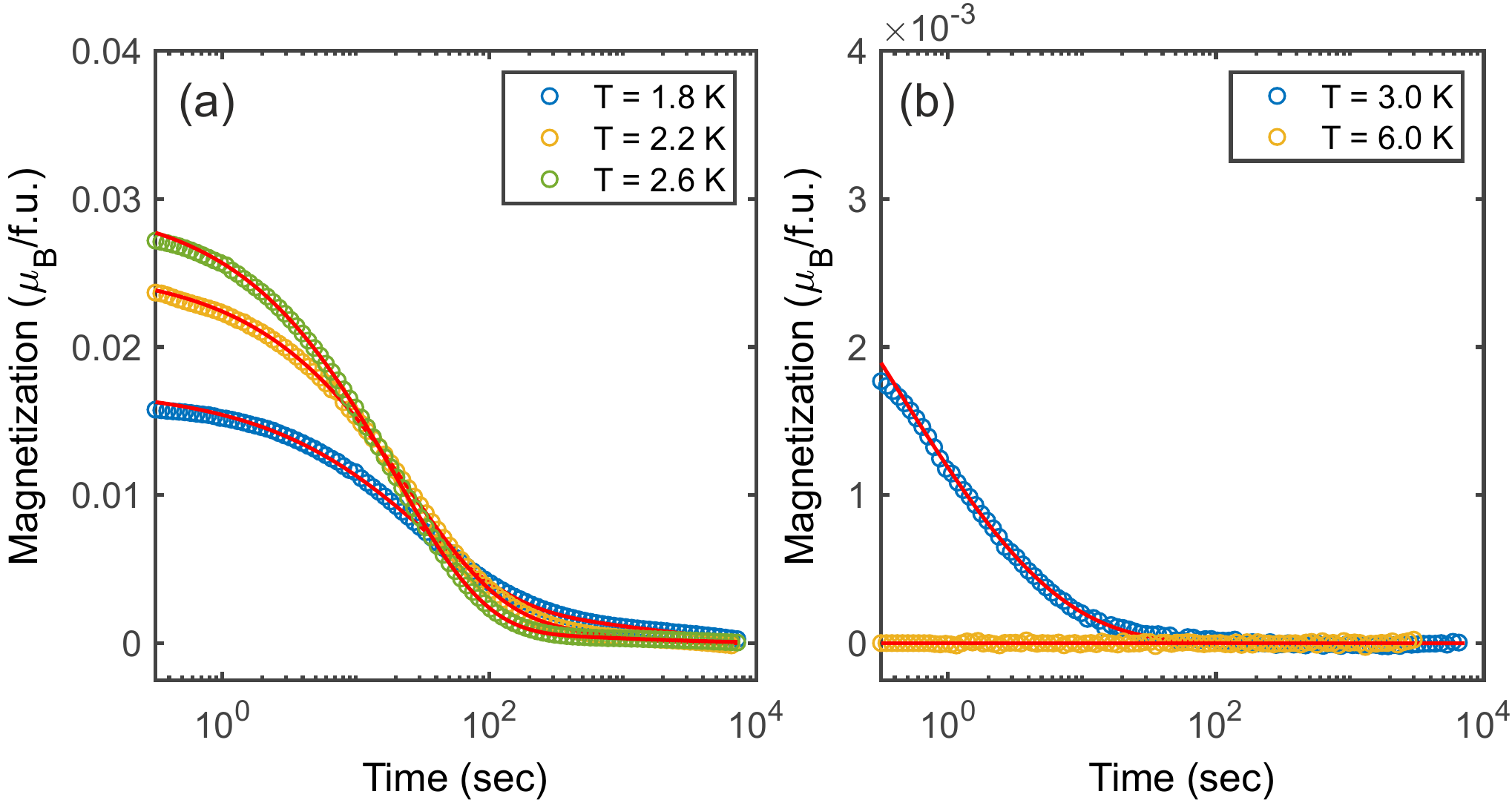}
  \caption{~Time dependence of magnetization taken after switching off 0.01~T magnetic field. Panels (a) and (b) show low- and high-temperature data. Note that $y$-scale is different in (a) and (b).}
  \label{Fig:relax}
\end{figure}

To characterize the timescale of the magnetization dynamics below \TN\ we used a VSM magnetometer and measured magnetization relaxation. To do this, we applied the following protocol: (i) ZFC to the base temperature $T = 1.8$~K; (ii) apply 0.01~T external field; (iii) wait for 300~s; (iv) decrease the external field to zero with sweep rate of 0.07~T/s; (v) collect time-dependent $M(t)$ for 3~hours; (vi) increase temperature to the new target $T$ and repeat from the step (ii). The relaxation curves collected at several selected temperatures above and below \TN\ are shown in Fig.~\ref{Fig:relax}.

Conventionally, the relaxation process $M(t)$ can be described with an exponential function, but we were not able to obtain good fit of our experimental data using a single exponent at $T < 3$~K. The measured curves were fitted with a sum of two exponential functions:
\begin{equation}
    M(t) = M_1 e^{-(t/\tau_1)^{\beta_1}} + M_2 e^{-(t/\tau_2)^{\beta_2}} 
    \label{eq:relax}
\end{equation}
where, $M_1$ and $M_2$ correspond to two relaxing moments, $\tau_1, \tau_2$ are the relaxation times, $\beta_1, \beta_2$ are stretching parameters.
To improve the fit quality we used $\beta_1$ and $\beta_2$ as global parameters for fitting of $T = 1.8, 2.2$ and 2.6~K curves [Fig.~\ref{Fig:relax}~(a)].
The fitted curves are shown by solid lines in Fig.~\ref{Fig:relax}. Results of the fitting yield $\beta_1 = 0.35$ and $\beta_2 = 0.6$ and $\tau_1 = 550$~s, $\tau_2 = 25$~s at $T = 1.8$~K; the relaxation times exhibit only a minor increase with temperature up to 2.6~K. 
The relaxation curve taken at $T = 3$~K can be described by a single exponent with $\beta = 0.3$ and $\tau = 0.2$~s [Fig.~\ref{Fig:relax}~(b)].

\begin{figure}
  \includegraphics[width=1.0\linewidth]{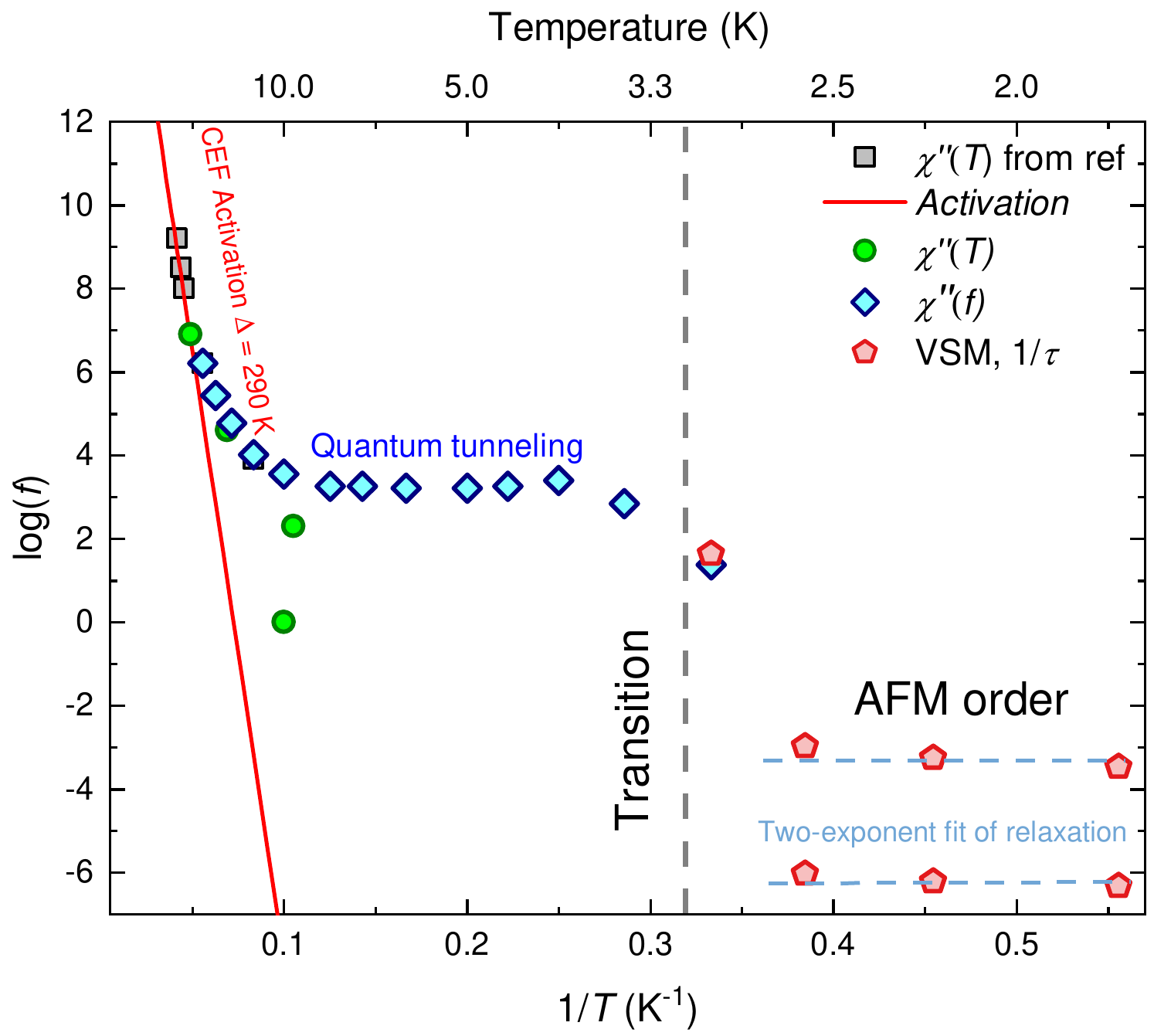}
  \caption{~Arrhenius plot log$(f)(1/T)$, reconstructed from our results along with data from Ref.~\cite{ke2009low}. The grey dotted line shows the transition temperature \TN. The blue and green points show the peak positions extracted from the frequency and temperature dependencies of imaginary part of AC susceptibility $\chi''$, and the black squares show the high-frequency data from Ref.~\cite{ke2009low}. Red solid line shows the calculated CEF activation curve with experimentally determined $\Delta = 290$~K.
  Red points show inverse relaxation constants $1/\tau$ extracted from a fit of the magnetic relaxation curves with Eq.~\eqref{eq:relax}.}
  \label{Fig:Arrhenius}
\end{figure}

The temperature dependence of observed relaxation times extracted from AC susceptibility and magnetization relaxation measurements is summarized in Fig.~\ref{Fig:Arrhenius}. Informed by our own INS measurements~\cite{wu2017magnetic}, we highlight three different regimes: (i) Arrhenius regime at high-temperature, $T > 10$~K; (ii) temperature-independent relaxation between 10~K and \TN; (iii) slow relaxation in the AFM phase. The temperature dependent AC-susceptibility in \dso\ was also studied in Ref.~\cite{ke2009low}, where the authors observed peak-like anomalies in $\chi''(T)$ curves, which shifted with frequency. The authors associated this peak with an Arrhenius-like relaxation process, taking the population of the CEF level into account, and extracted $\Delta/k_{\mathrm{B}} = 229$~K.  Our own later INS measurements indicated that the first CEF level is located at higher energy, $\Delta/k_{\mathrm{B}} = 290$~K~\cite{wu2017magnetic}. The calculated curve for a 290~K gap is shown in Fig.~\ref{Fig:Arrhenius} by a red line, and one can see good agreement with experimental points at high temperature and a clear crossover between regimes (i) and (ii).

The behavior seen in \dso\ strongly resembles the slowing down of the spin dynamics in a classical spin-ice compound \dto, which also shows three regimes at different temperatures: (i) Arrhenius relaxation at $T > 15$~K; (ii) plateau at $1.5 < T < 15$~K; (iii) another Arrhenius regime below $1.5$~K due to development of the spin-ice regime~\cite{matsuhira2001novel,matsuhira2011spin, takatsu2013ac}. The difference in the low-temperature regime (iii) is a consequence of the different ground states, AFM in \dso\ vs. spin-ice in \dto.

\subsection{Field-induced anomalies}

\begin{figure}[tb]
  \includegraphics[width=1.0\linewidth]{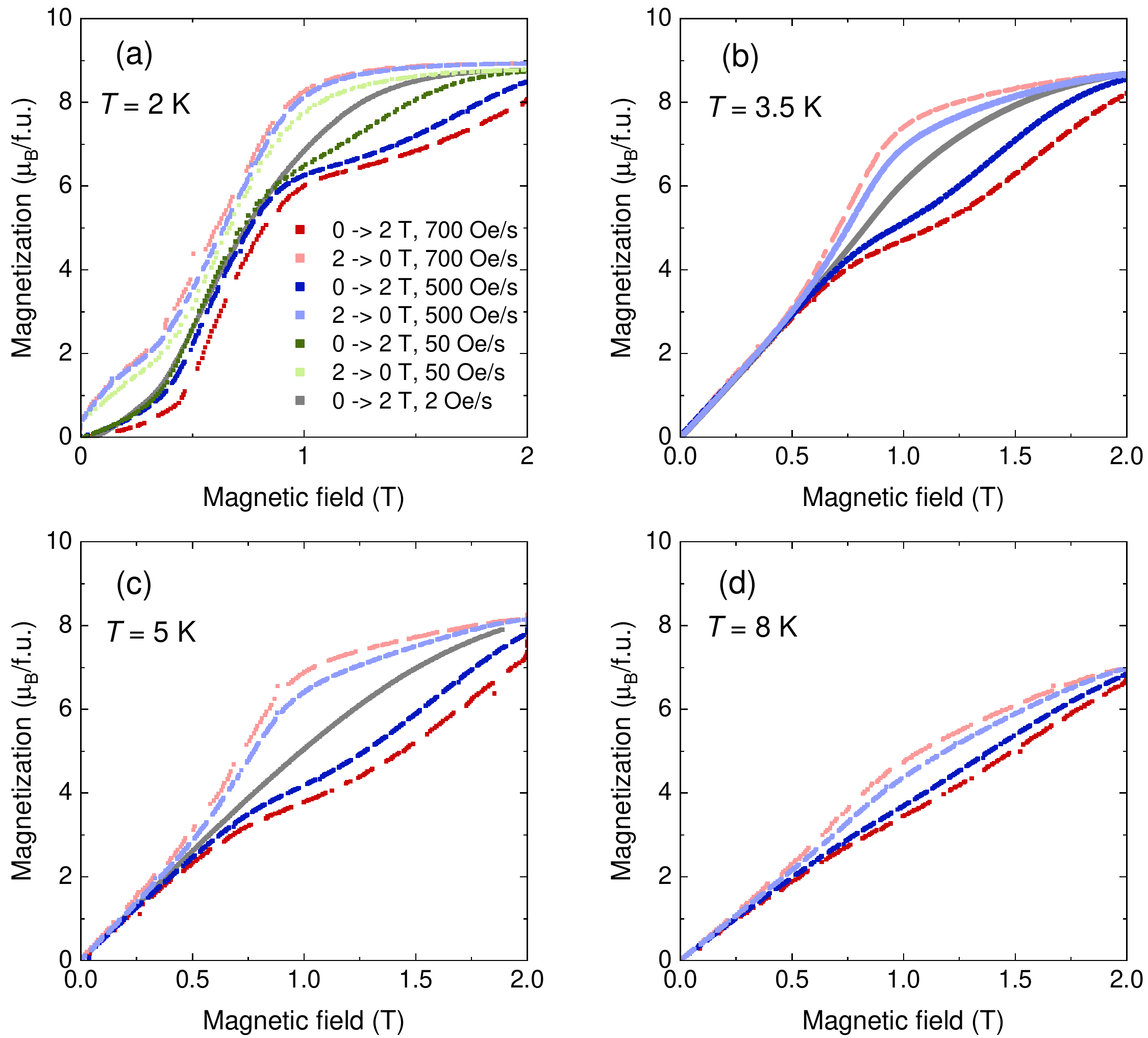}
  \caption{~Magnetization curves $M(B)$ measured at several temperatures as indicated at each panel. The curves were measured with different sweep field rates as shown in legend.}
  \label{Fig:MvsB}
\end{figure}

We next proceed with a description of the field-induced physics in \dso.  Authors of~\cite{wu2017magnetic} reported magnetization curves measured at $T = 2$~K along three crystallographic directions. Interestingly, the curves measured along $[010]$ and $[100]$-axes show two consecutive hystereses at low field and just below the saturation. These features were interpreted as two field-induced first-order phase transitions. Motivated by those observations we measured the magnetization of \dso\ at several temperatures with magnetic field applied along the $[010]$ axis. Figure~\ref{Fig:MvsB} shows magnetization curves collected at several temperatures below and above \TN. One can see that at $T = 2$~K magnetization curves measured with high sweep rates (50, 500 and 700~Oe/s) show considerable hysteresis over the whole field range. However, we can clearly highlight two distinct transitions: the first kink at $B \approx 0.4$~T and the second anomaly at $B \approx 1$~T. Noticeably, when the sweep rate decreases, only the low-field features remains visible, while magnetization at higher fields shows simple Brillouin-like behavior. Figures~\ref{Fig:MvsB}(b-d) demonstrate magnetization collected above \TN\ and one can see that the magnetization is perfectly linear at the low-field regime in agreement for the expectation for a paramagnet. However, the clear kink as well as the hysteresis at 0.7--2~T are clearly seen for high field-sweep rates, while the low-sweep curves show simple Brillouin-like behavior.

Based on these data we can associate the first transition with the field-induced destruction of the AFM order, while the high-field hysteresis is associated with single-ion physics, because it persists to temperatures up to $\sim 8$~K, which is much larger than the characteristic energy scale of magnetic interactions in \dso. We associate this effect with the strong magnetocaloric effect (MCE) in \dso~\cite{wu2019antiferromagnetic} and in the next section we show that Monte-Carlo simulations, which take into account the MCE, are capable to reproduce this effect.

\subsection{Monte-Carlo simulations}
\label{SubSec:MCS}

As discussed above, \dso\ shows strong Ising-like single-ion anisotropy of magnetic moments, meaning that Dy moments can be pointed up or down along the CEF-dictated easy magnetization direction at each site. In addition, a previous report associated magnetic order with the dipole-dipole interactions between Dy moments~\cite{wu2017magnetic}. The standard approach to describe physical properties of an Ising system on a 3D lattice is by Monte-Carlo modeling and here we make use of the metropolis algorithm to describe magnetic behavior of \dso~\cite{beichl2000metropolis}. We considered a $10\times10\times10\times4$ cluster of Ising spins, which are coupled by dipole-dipole interaction and are in thermal contact with a reservoir at temperature $T_{\mathrm{res}}$. Most parameters of our model, such as interatomic distances, the field- and temperature dependence of the thermal conductivity (approximated from the data measured on isostructural DyAlO$_3$), the magnetic moment of Dy$^{3+}$ and the direction of easy axis were fixed from the experimental data~\cite{wu2017magnetic, numazawa1998thermal}. The only free parameter is the coefficient which converts the number of Monte-Carlo steps to the experimental time, which we fixed by comparison of calculated and experimental $\chi''(f)$ curves and the absolute value of the thermal conductivity. See Sec.~\ref{Sec:MC} for details. 

\begin{figure}
  \includegraphics[width=1.0\linewidth]{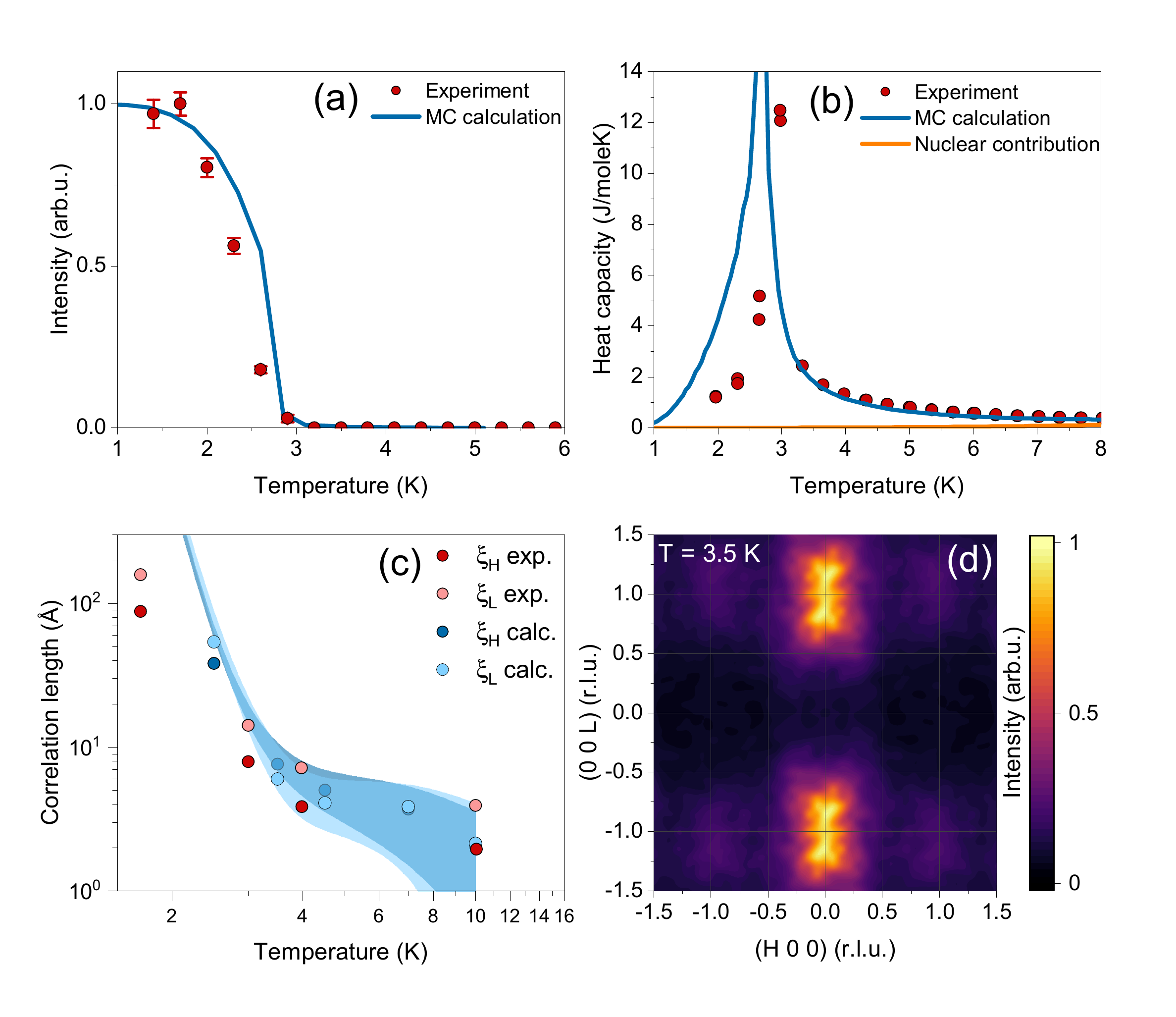}
  \caption{~AFM ordering at zero field. 
  (a)~Temperature dependence of (001) Bragg peak. Red dots represent experimental data for square of the ordered moment~\cite{wu2017magnetic} and solid line is the result of Monte-Carlo calculations.
  (b)~Calculated and measured low-temperature specific heat. The nuclear contribution was approximated by $C(T) = \beta T^3$ with $\beta = 0.00023$~J/mole$\cdot$K$^{4}$ and added to the calculated curve. The nuclear contribution reaches 0.047~J/mole K at $T = 6$~K and therefore is barely visible at this temperature scale.
  (c)~Temperature dependence of the correlation length measured by neutron diffuse scattering (red and pink dots) and calculated by Monte Carlo (blue and light blue dots). The filled area represents the estimated uncertainty of the calculation.
  (d)~Simulated neutron diffraction pattern in the $(H0L)$ scattering plane calculated at $T_{\mathrm{N}} < T = 3.5$~K showing anisotropic diffuse scattering around the primary AFM peak (001).}
  \label{Fig:transition}
\end{figure}

We start the presentation of our calculations with the low-temperature magnetic structure. First, our data show that the $GxAy$ is indeed the ground state of \dso\ and has the lowest energy among the four possible $\mathbf{k} = 0$ magnetic configurations in \dso, in agreement with previous estimates for smaller  clusters~\cite{wu2017magnetic, wu2019antiferromagnetic}. As the next step we calculated the static magnetic structure factor, $S(\mathbf{Q})$, and found that the AFM order manifests itself with a strong magnetic Bragg peak at $\mathbf{Q} = (001)$. We plot the calculated temperature dependence of (001) peak in Fig.~\ref{Fig:transition} along with the experimental measurements of the ordered moment~\cite{wu2017magnetic}. Fitting of calculated data near the critical temperature was performed using the following expression:

\[
    S = \left\{\begin{array}{lr}
    0, & \text{for } (T > T_c)\\
    S_0 (1 - \big( \frac{T}{T_c} \big)^{\beta})^{\alpha}, & \text{for } (T \leq T_c),
    \end{array}\right\}
\]
and yielded the critical temperature  $T_{\mathrm{N}} = 2.853(1)$~K. The very good  agreement with the experimentally determined $T_{\mathrm{N}} = 3.11$~K indicates that the dipolar interaction is the primary magnetic interaction in \dso. We have also modeled the magnetic diffuse scattering above the transition temperature. The representative diffuse pattern calculated at $T = 3.5$~K is represented in Fig.~\ref{Fig:transition}~(d), which was obtained from averaging a number of Monte-Carlo runs. 
We extracted the temperature dependence of the correlation length [Fig.~\ref{Fig:transition}~(c)] and compared it with experimental results measured at the CNCS instrument~\cite{wu2017magnetic}. Clearly, the measured and the calculated curves show good agreement with $\xi_c >\xi_{ab}$ over all temperature range. The reason is the mutual arrangement of crystal axes and directions of Dy$^{3+}$ moments: the nearest neighbor Dy moments along the $c$ axis have strong antiferromagnetic interaction, while interaction between in-plane Dy moments is nearly canceled out for the $GxAy$ magnetic configuration~\cite{wu2019antiferromagnetic}. To further characterize the magnetic transition we calculated the specific heat of \dso\ and show it along with the experimental curve in Fig.~\ref{Fig:transition} (b). Both curves show a sharp $\lambda$-like peak at \TN\ due to AFM transition with only weak shoulders above \TN, which indicates weak magnetic correlations in the paramagnetic state. The entropy change associated with anomaly in the specific heat is in agreement with the value expected for a Ising system: $R\ln(2S + 1)$ for $S = 1/2$.

To reveal the origin of the bifurcation between the $M(T)$ curves shown in Fig.~\ref{M_vs_T_rate}~(a) we performed Monte Carlo simulations of the temperature dependences of magnetization [see Fig.~\ref{M_vs_T_rate}~(b)] with different sweep rates of the reservoir temperature, $dT_{\mathrm{res}}/dt$. We found that the calculated magnetization curves are in qualitative agreement with experimental data. Specifically, the magnetization curve calculated for $dT/dt = 0.1$~K/min shows a clear cusp at \TN\ and a weak hysteresis below the transition. With increase sweep rate $dT/dt$ the hysteresis becomes wider and the peak for warming up curves shifts towards higher temperatures in good qualitative agreement with the experiment. The reason for this behavior is the poor thermal stabilization due to both, weak thermal conductivity and slow thermalization during the temperature sweeps, because of which the actual temperature of the sample can differ considerably from $T_{\mathrm{res}}$. This effect causes bifurcation between cooling and warming $M(T)$ curves and the sweep-rate dependence of the transition temperature.

\begin{figure}
  \includegraphics[width=1\linewidth]{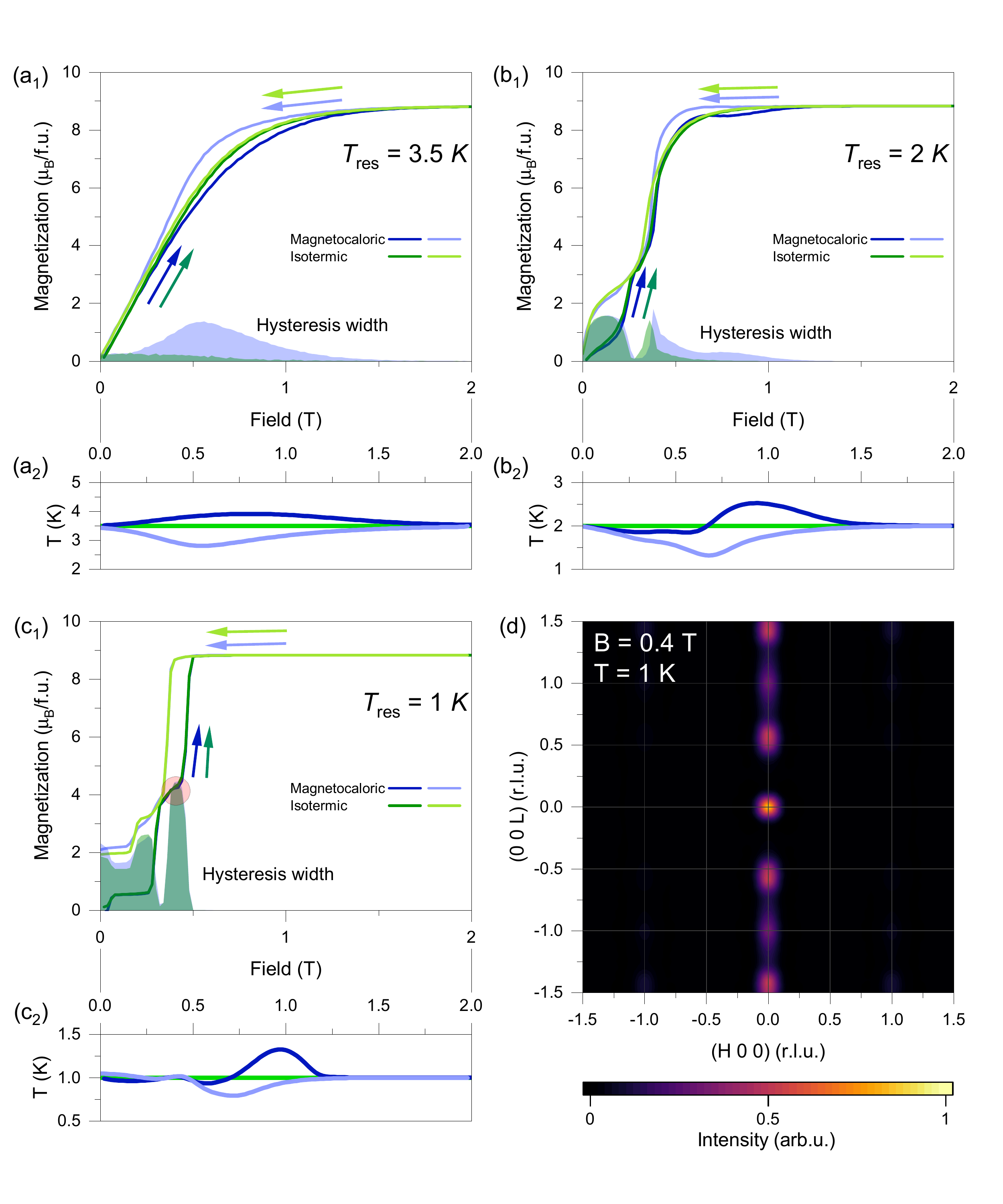}
  \caption{~(a1-c1)~The magnetization curves calculated taking into account magnetocaloric effect at various temperatures of reservoir and corresponding system temperature dependence. The magnetization curves were calculated for external field ramped from zero to 2~T and backward, the shaded areas show width of the hysteresis curves $\Delta{}M(B) =  M(B)\uparrow - M(B)\downarrow$. Additional  panels~(a2-c2) show temperature change during simulation, grey horizontal lines denote reservoir temperatures. Reservoir temperatures are indicated in each panel. The contour map (d) is the calculated magnetic structural factor for the incommensurate state around $B = 0.4$~T and $T = 1$~K (red circle point at (c1)).}
  \label{magcal}
\end{figure}

As the next step we calculated the magnetic field dependence of the magnetization at three characteristic temperatures of the reservoir, $T_{\mathrm{res}} = 1,~2$ and 3.5~K and the results are summarized in Fig.~\ref{magcal}. We performed two versions of calculations, (i) isothermal magnetization with $T_{\mathrm{Sample}} = T_{\mathrm{res}}$ strictly maintained, and (ii) by taking into account the strong magnetocaloric effect~\cite{wu2019large}, see details of calculations in Sec.~\ref{app:MC}. The field sweep rate for all curves in Fig.~\ref{magcal} was $0.5$~K/min.  

First of all, let us consider the magnetization curves calculated at $T = 3.5$~K above the ordering temperature, which are shown in Fig.~\ref{magcal}~(a1). The isothermal magnetization calculated at $T = 3.5$~K shows featureless Brillouin-like behavior in agreement with expectations for a paramagnet. Introduction of the magnetocaloric effect changes qualitative behavior of the magnetization curves and opens a broad hysteresis at $B \approx$ 0.4--1.2~T, because of the field-induced change of $T_{\mathrm{samp}}$, which is shown in Fig.~\ref{magcal}~(a2). Thus we conclude that the hysteresis at intermediate fields observed in our magnetization measurements for $dB/dt \geq\ 50$~Oe/s (Fig.~\ref{Fig:MvsB}) is caused by the magnetocaloric effect, which however plays a minor role when the field ramp rate is small compared to the thermalization time, meaning that $T_{\mathrm{samp}} = T_{\mathrm{res}}$. 

Below the magnetic transition temperature the magnetization curve changes considerably. The curves calculated at $T_{\mathrm{res}} = 2$~K demonstrate a metamagnetic transition and a broad hysteresis at low fields, $B < 0.3$~T, which behaves similarly for both, the isothermal and non-isothermal curves [Fig.~\ref{magcal}~(b1)]. Above the metamagnetic transition the isothermal curves merge and show the featureless Brillouin-like behavior, while a narrow hysteresis emerges at $B = 0.5$--1~T for the curves calculated with magnetocaloric effect. We note that the exact shape of the calculated curve depends considerably on the thermal conductivity of \dso, which depends on both magnetic field and temperature, $\lambda(B,T)$. This quantity was not measured in the low-temperature regime, $T < 10$~K and in our calculations we make use of the data measured in the paramagnetic phase on the isostructural DyAlO$_3$~\cite{numazawa1998thermal}, which we approximate down to the temperature range of interest~(see Sec.~\ref{app:MC} for details). This procedure does not provide quantitatively precise values for the thermal conductivity, especially below \TN\ where thermal conductivity can exhibit strong anomalies near the critical field~\cite{metcalfe1972magnetothermal, zhao2012heat, dixon1980low}.  Therefore, the exact shape of calculated and observed magnetization curves does not match quantitatively. However, our calculations allow us to associate the high-field hysteresis with the magnetocaloric effect, while the low-field anomalies, including the metamagnetic transition and the hysteresis of magnetization, are associated with collective behavior.

Figure~\ref{magcal}~(c1) shows magnetization calculated at $T_{\mathrm{res}} = 1$~K and one can see that isothermal and non-isothermal curves coincide over all field scales, meaning that the magnetocaloric effect has a minor influence on the magnetization curve at this temperature. Moreover, the curves calculated for $B = 0 \rightarrow\ 2$~T demonstrate a narrow $M_{\mathrm{s}}/2$ plateau at $B = 0.4$~T. 
In a system with Ising-like anisotropy, the magnetic moments can point parallel or antiparallel to the easy-axis direction, therefore below the ordering temperature, one can expect to see the formation of SDW-like incommensurate phases at the intermediate magnetic fields~\cite{fisher1980infinitely, selke1988annni}. Figure~\ref{magcal}~(d) shows the calculated static spin structure factor at $T = 1$~K and $B = 0.4$~T, which corresponds to the plateau field range. One can see clearly formation of the incommensurate peaks at $\mathbf{q} = (0~0~0.55)$, which can be associated with the formation of $\uparrow\uparrow\uparrow\downarrow$ type of order along the $c$ axis. Two other reflections correspond to the primary magnetic reflection of the zero-field AFM phase, $\mathbf{q} = (0~0~1)$ and ferromagnetic $\mathbf{q} = (0~0~0)$ peak. We note that a similar field-induced incommensurate magnetic order with $\mathbf{k} = (0~0~1\pm\delta)$ was observed previously in the isostructural material YbAlO$_3$~\cite{Wu2019Tomonaga, nikitin2021multiple}.  Low-temperature in-field neutron diffraction measurements would be required to verify the presence of the incommensurate magnetic phases in \dso.

\section{Discussion and Conclusion}

We demonstrate that \dso\ exhibits slow dynamics of magnetization below $T^* = 10$~K, which is caused by strong CEF-induced uniaxial anisotropy of Dy$^{3+}$ moments. 
We note that similar behavior was observed previously in multiple organic-based magnetic molecules~\cite{guo2011relaxation, langley2014modulation, gatteschi2003quantum}, while such a behavior in inorganic materials remains relatively rare. Several representative examples are \dto\ and Ca$_3$Co$_2$O$_6$. However, in addition to the QTM, both aforementioned materials display complex magnetic behavior due to frustration of magnetic interactions. The presence of the competing interactions makes it difficult to disentangle the effects of QTM and magnetic frustration on spin freezing in these systems, contrary to the case of \dso. We have characterized the low-temperature spin dynamics of \dso\ using AC susceptibility and magnetization relaxation measurements and have observed three regimes: Arrhenius-like behavior at $T > T^*$, a plateau between $T^*$ and \TN, and slow relaxation of magnetization below \TN. Thus, \dso\ represents a so-far unique example of a classical dipolar AFM, which combines a classical magnetically-ordered ground state with QTM. 

Our Monte Carlo simulations appear to capture the essential physics of \dso, including the magnetic ground state, the temperature of the AFM transition, slow dynamics of magnetization at low temperatures and bifurcation between $M(T)$ curves collected upon cooling and warming. In addition, by taking into account the magnetocaloric effect we were able to describe the magnetization curves and demonstrate that the hysteresis in the paramagnetic phase is caused by considerable field-induced change of the sample temperature during the measurements. Our simulations also predict the formation of an incommensurate spin-density wave magnetic phase at low temperatures and intermediate magnetic fields, similar to that observed in YbAlO$_3$~\cite{Wu2019Tomonaga, nikitin2021multiple}, whose existence in \dso\ awaits experimental verification with elastic neutron scattering measurements.

We note that although our model captures the main features of magnetic behavior in \dso\ there are  minor quantitative disagreements such as the exact shape of the hysteresis curves shown in Figs.~\ref{Fig:MvsB}~(a-c), which could probably be improved by including in the model exact results for the field- and temperature-dependence of thermal conductivity. In addition, the $\chi''(\chi')$ curve has an asymmetric shape indicating a complex distribution of the relaxation times, while our model implies a single relaxation channel for simplicity. 

To conclude, we have applied AC susceptibility and DC magnetization measurements, supported by specific heat, neutron diffuse scattering and Monte-Carlo calculations to characterize spin dynamics in \dso. Our results indicate that \dso\ represents a rare combination of single-ion QTM behavior with classical dipolar interactions and stimulate further search of rare-earth based condensed matter systems with QTM.

\begin{acknowledgments}
{\it Acknowledgments.} We thank O. Stockert and A. S. Sukhanov for stimulating discussions. S.E.N. acknowledges financial support from the European Union Horizon 2020 research and innovation program under Marie Sklodowska-Curie Grant No.~884104. The work of N.D.A. was supported by the German Research Foundation (DFG) through grant No. PE~3318/3-1. This research used resources at the Spallation Neutron Source, a DOE Office of Science User Facility operated by the Oak Ridge National Laboratory.


\end{acknowledgments}

\appendix
\newpage
\section{Experimental details}\label{app:Exp}
High-quality single-crystals of \dso\ were obtained commercially~\cite{mtixtl}.
Magnetization and AC-susceptibility measurements were performed in a temperature range 1.8 - 100~K, using an MPMS SQUID VSM instrument by Quantum Design. For the magnetic measurements we used a sample with mass of 1.6~mg. Magnetic field was applied along the [010] direction, which is the easy axis of magnetization. Specific heat was measured using PPMS from Quantum Design. 
The single crystal neutron scattering measurements were done at the Cold Neutron Chopper Spectrometer (CNCS) \cite{CNCS1,CNCS2}. To quantify the correlation lengths $\xi_H$, $\xi_L$ we perform a two dimensional fitting of the (0~0~1) magnetic peak using the following function~\cite{wu2017magnetic}:

\begin{equation}
S_{\mathrm{mag}}(\mathbf{Q}) = \frac{\sinh{a/\xi_H}}{(\cosh{a/\xi_H} - \cos{\pi q_{H}})}
\frac{\sinh{c/\xi_L}}{(\cosh{c/\xi_L} - \cos{\pi q_{L}})},
\end{equation}
where $a$ and $c$ are the lattice parameters. The calculated magnetic structure factor (see Sec.~\ref{sec:diffraction}) was fitted using the same equation. 

\section{Monte-Carlo simulations}\label{app:MC}
\label{Sec:MC}
\subsection{Magnetic Hamiltonian}
At low temperature magnetic system of Dy$^{3+}$ ions can be effectively described by dipole-dipole interaction. Since the magnetic properties of \dso\ exhibit Ising-like behavior, it is convenient to theoretically study these with Monte Carlo (MC) simulations. In this work we performed MC simulations of a 3D Ising system of Dy$^{3+}$ moments using the classical Metropolis single-flip algorithm~\cite{beichl2000metropolis}. The Hamiltonian used in the calculations was taken as a combination of dipole-dipole interaction energy and a Zeeman term due to the external field. The Hamiltonian reads:

\begin{equation} 
\begin{array}{rcl}
\mathcal{H} & = & -\displaystyle\frac{1}{2}\frac{\mu_0}{4\pi}\sum_{i,j}\left[\frac{3(\vec{m_i},\vec{r_{i,j}})(\vec{m_j},\vec{r_{i,j}})}{|\vec{r_{i,j}}|^5} - \frac{(\vec{m_i},\vec{m_j})}{|\vec{r_{i,j}}|^3}\right] \\ [5ex]
  &  & - \displaystyle{B}\sum_{i}\vec{m_i} {\;},
\end{array}
\end{equation} 

where the first term is the dipole-dipole interaction energy and the second one is the energy of Zeeman interaction of the magnetic moments and external field. Because the summation goes over each interaction twofold there is the \(\frac{1}{2}\) factor in front of the first term. The $\mu_0$ corresponds to vacuum permeability constant, $\vec{m_i}$ is the magnetic moment on the $i$ site, $\vec{r_{i,j}}$ is the radius vector between the $i$ site and $j$ site magnetic moments, the $B$ is the external field. In the calculations the magnitude of magnetic moments was taken as $10\mu_B$.

\begin{figure}
  \includegraphics[width=0.7\linewidth]{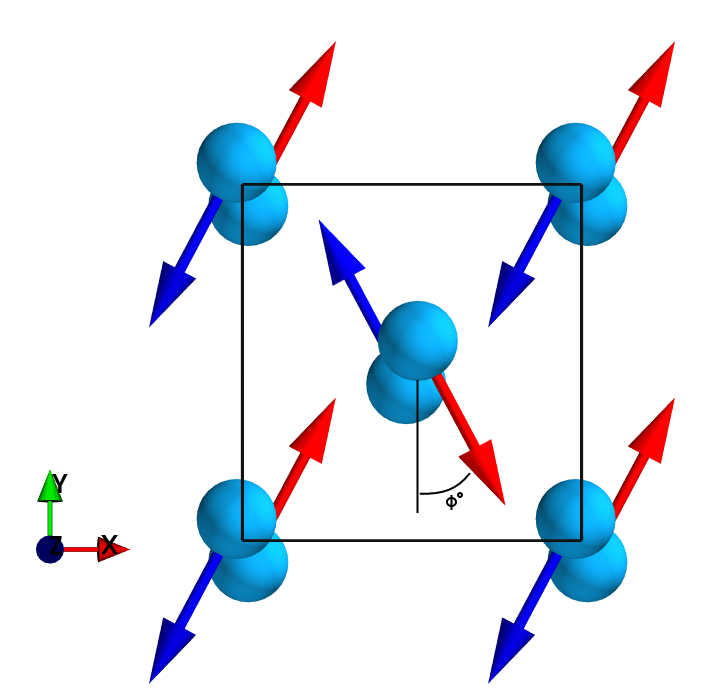}
  \caption{~Scheme of GxAy spin configuration.}
  \label{str_GxAy}
\end{figure}

The crystal structure of the \dso\ was taken into account during the simulations by implementing the 3D Ising system with inter-atomic distances corresponding to the crystallographic data. This is necessary because inter-atomic radius vectors appear in the expression for the dipolar energy. The size of the system in the simulations was $10\times10\times10$ unit cells each containing 4 spins (4000 spins in total). The dipole-dipole interaction was calculated for up 62 interactions per one site (the cutoff distance between neighbours was 9.5~\AA) as a compromise between calculation time and accuracy.  A test simulation showed that an inclusion of further neighbors produced minor effects on the magnetic behavior. The periodic boundary conditions were used. Each Ising spin can be in the $s = 1$ or $s = -1$ state corresponding to the spin vector $\vec{S} = s(\pm\sin(\phi), \cos(\phi), 0)$, where the sign in front of the $\sin$ term depends on the position of site in unit cell (see Fig.~\ref{str_GxAy}). $\phi$ is the angle between spin vector and the $[0 1 0]$ or the $[0 \bar{1} 0]$ direction determined by CEF and is equal to 28$^{\circ}$~\cite{wu2017magnetic}.

\subsection{AC-susceptibility}
\begin{figure}
  \includegraphics[width=1\linewidth]{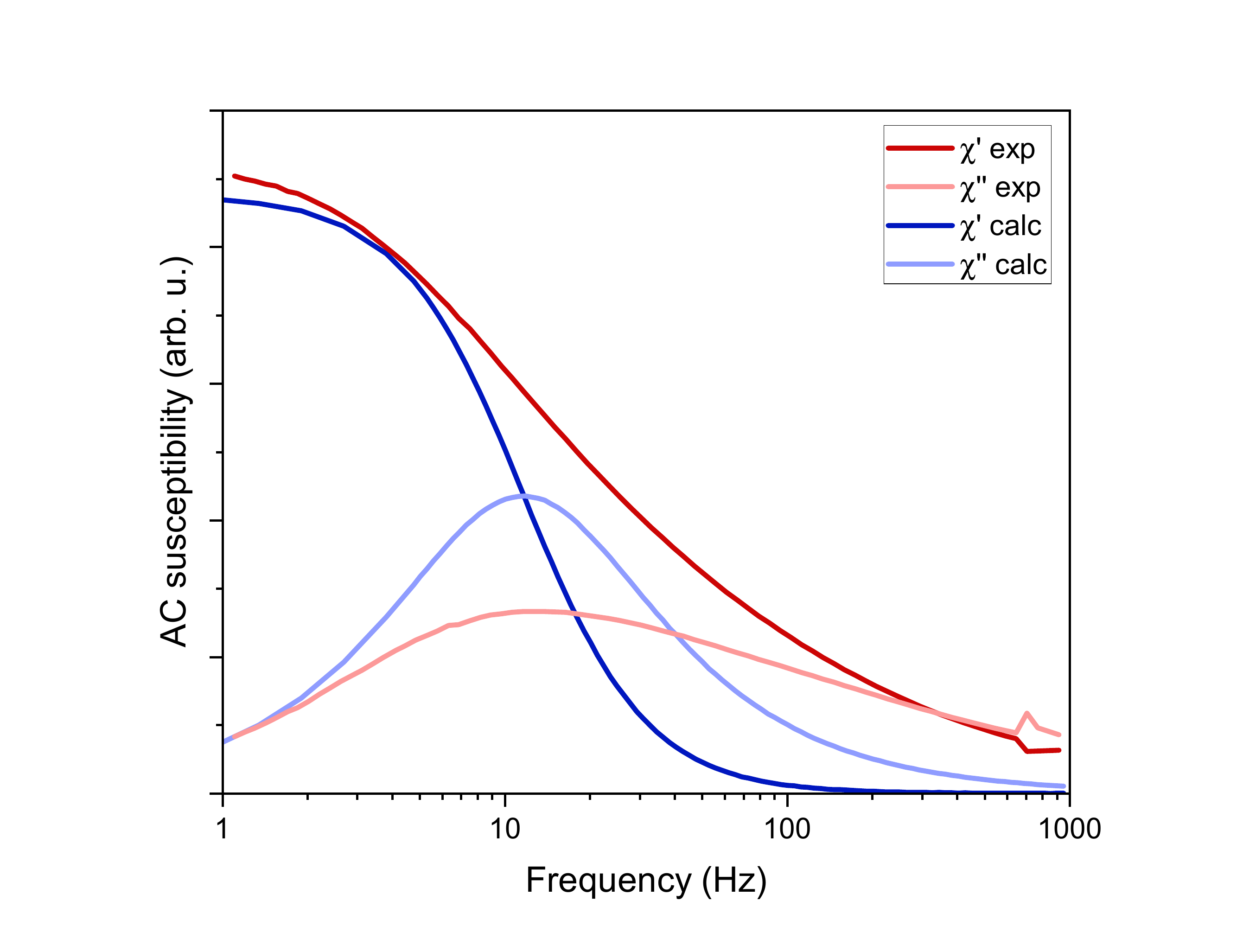}
  \caption{~Calculated and experimental AC susceptibility at $T = 4$~K.}
  \label{susc_comp}
\end{figure}

AC susceptibility data were used in our simulations to adjust the time-scale transformation coefficient. For calculations of real and imaginary parts of AC susceptibility the following expressions were used:

\begin{eqnarray}
\chi'(f) & = & \frac{1}{B_0 N_{\mathrm{total}}}\sum_{i = 0}^{N_{\mathrm{total}}} M_i\sin(2{\pi}f\frac{i}{A} - \frac{\pi}{2}) \\ [1ex]
\chi''(f) & = & \frac{1}{B_0 N_{\mathrm{total}}}\sum_{i = 0}^{N_{\mathrm{total}}} M_i\cos(2{\pi}f\frac{i}{A} - \frac{\pi}{2}) {\;},
\end{eqnarray}

where $\chi'$ and $\chi''$ are real and imaginary parts of AC susceptibility, $B_0$ and $f$ are the magnitude and the frequency of the alternating external field, and  $M_i$ is the magnetization of the system at $i$-th MC step. The alternating external field at the $i$-th MC step is taken as 
\(B = B_0\sin(2{\pi}f{i/A} - \pi/2)\). Used external field amplitude $B_0$ is $0.1~T$.
The $A$ parameter is the conversion factor between real time and simulation time \(t_{sim} = At_{real}\). In the above expressions, an averaging was performed over numerous periods, with $N_{total}$ up to $10^6$ MC steps per spin. In Fig.~\ref{susc_comp} experimental and calculated AC susceptibility at temperature 4~K were drawn on the same layer for comparison. The calculated imaginary part of susceptibility $\chi''$ consists of one symmetrical peak associated with a particular relaxation time. In contrast, the experimental dependence of $\chi''$ is more broad and asymmetric. As it was discussed in Sec.~\ref{app:SDMO}, the shape of experimental $\chi''$ curves can be interpreted as a presence of more then one relaxation channel. The value of the $A$ parameter was adjusted for fitting the calculated position of $\chi''$ maximum to the experimental value and later the $A$ parameter was fixed when used in further calculations.

\subsection{Magnetic diffraction} \label{sec:diffraction}
The magnetic structure factor was calculated as a spatial Fourier transform of the spin-spin correlation function taking into account the polarization factor of neutron scattering using the following expression:

\begin{equation}
S_{mag}(\vec{q}) = \sum_{\alpha,\beta}\big[ \{\delta_{\alpha,\beta} - \frac{q_{\alpha}q_{\beta}}{|\vec{q}|^2}\}\sum_{j, j'}S_{j, \alpha}S_{j', \beta}\exp(i\vec{q}(\vec{R_j} - \vec{R_{j'}}))\big],
\end{equation}
where $\vec{q}$ is the wave vector, $\alpha$ and $\beta$ are Cartesian components - $x, y$ and $z$, the $j$ and $j'$ are the variables which go over all the sites in system, the $S_{j,\alpha}$ is the $\alpha$ component of spin vector on $j$-th site. The term in front of the second summation is the polarisation factor. The isotropic magnetic form factor of Dy ions, which gives a monotonic suppression of the intensity at large $Q$, was not included to the calculations.



\subsection{Thermodynamic properties}

\begin{figure}
  \includegraphics[width=1\linewidth]{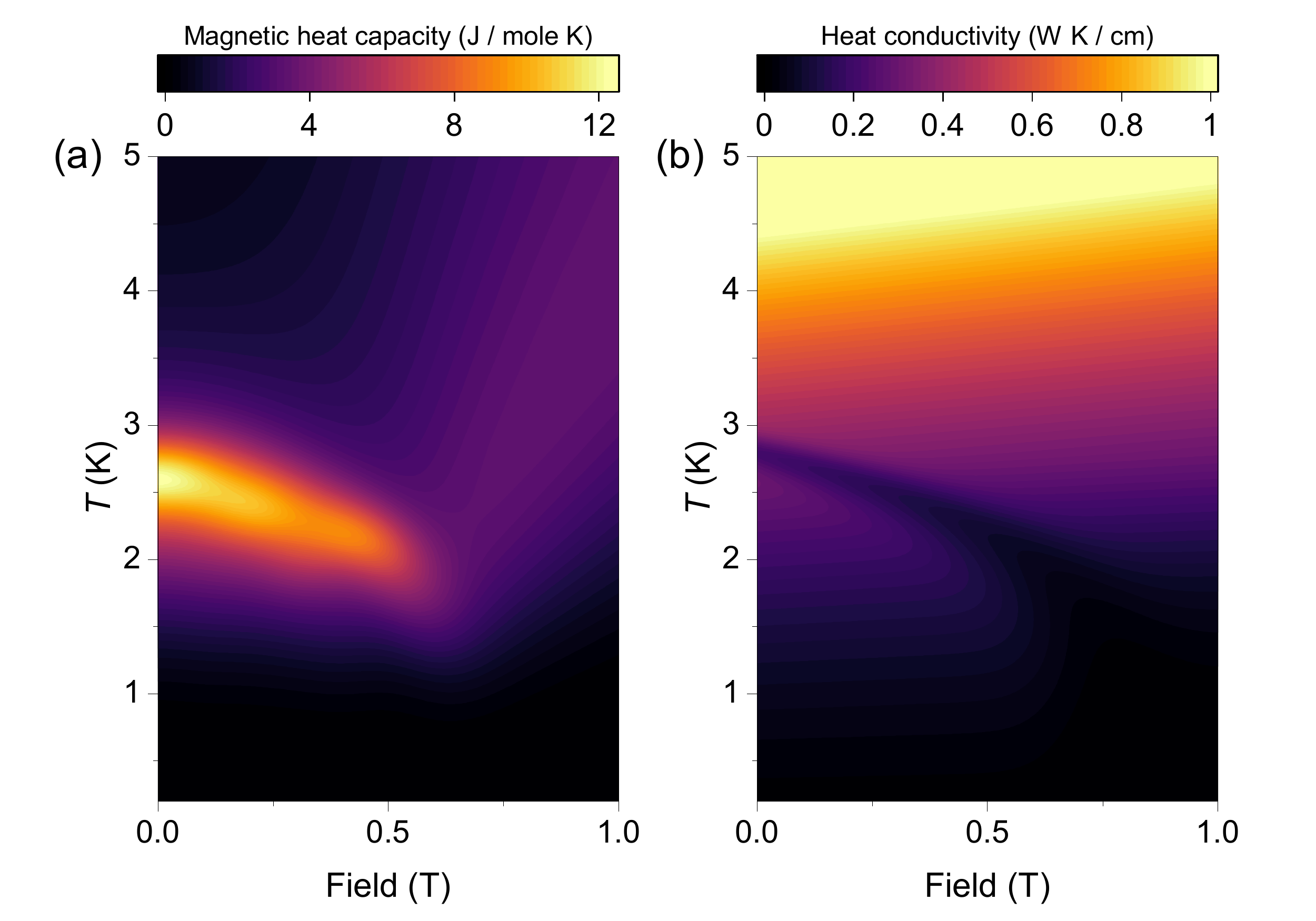}
  \caption{~(a) Calculated magnetic part of the heat capacity as function of external field and temperature. (b) The heat conductivity extrapolated in nonzero external field region.}
  \label{hchc}
\end{figure}

In order to achieve a better agreement with experimental measurements the magnetocaloric effect was taken into account for calculations of the field dependence of magnetization. The heat capacity used in the magnetocaloric calculations was the combination of the lattice contribution, which was obtained from approximation of experimental measurements, and the magnetic contribution which was calculated from energy fluctuations:

\begin{equation}
C_{\mathrm{mag}}(T, B) = \frac{{\langle \mathcal{H}^2 \rangle} - {\langle \mathcal{H} \rangle}^2}{k_{B}T^2},
\end{equation}

where $k_B$ is the Boltzmann constant and the angle brackets means averaging over numerous MC steps. The total heat capacity can be written as ${C}_{p} = C_{\mathrm{mag}} + C_{\mathrm{lat}}$, the lattice part was approximated from experimental data above ordering temperature \TN.
The magnetocaloric effect was implemented in calculations of the field dependence of magnetization as a variable system temperature. The temperature change comes form the magnetocaloric effect itself and also from thermal contact with the temperature reservoir. If the external field changes from $B_1$ to $B_2$ during time ${\Delta}t$ the system temperature change $\Delta T$ can be written as follows:

\begin{multline}
{\Delta}T = \frac{T}{\frac{1}{2}(C_p(B_1,T) + C_p(B_2,T))}\frac{1}{2}\bigg[\frac{{\partial}M(B_1,T)}{{\partial}T} + \\
+ \frac{{\partial}M(B_2,T)}{{\partial}T}\bigg]dB - \frac{T - T_{res}}{\frac{1}{2}(C_p(B_1,T) + C_p(B_2,T))}\times \\ \times\frac{1}{2}(\lambda(B_1,T) + \lambda(B_2,T)){\Delta}t,
\end{multline}

where the first term is the magnetocalorics and the second expresses thermal contact with reservoir. $dB = B_2 - B_1$ is the change of external field, $T$ is the system temperature before the correction, $T_{res}$ is the reservoir temperature, $\lambda(B,T)$ is the thermal conductivity coefficient. The field change $dB$ is assumed to be small enough in order to justify the replacement of the integral from original thermodynamic formula by this simple expression. The heat capacity $C_p(B,T)$ values were obtained as it was mentioned above. The time $\Delta t$ was recalculated into Monte Carlo steps with help of previously discussed $A$ parameter. The partial derivative $\frac{{\partial}M(B,T)}{{\partial}T}$ was calculated beforehand from equilibrium values of magnetization:

\begin{equation}
\frac{{\partial}M(B,T)}{{\partial}T} = \frac{\langle M(B, T + \Delta T)\rangle - \langle M(B, T - \Delta T)\rangle}{2\Delta T}.
\end{equation}

In order to numerically estimate the derivative the averaging was done over more $10^6$ MC steps per spin.

The exact shape of the field dependence of thermodynamic properties can significantly affect the magnetization behavior in magnetocaloric calculations. Since the experimental data for the heat conductivity $\lambda(B,T)$ is unknown in presence of external field and in low temperature for \dso, we had to use values obtained from extrapolation process and turn into account assumptions. The heat conductivity $\lambda(B,T)$ values in zero field were taken from data for the isostructural DyAlO$_3$ material. We assumed the following when extrapolating the values to nonzero field: (i) the heat conductivity $\lambda(B,T)$ slowly decreases in the external field~\cite{numazawa1998thermal}, and (ii) near the region of metamagnetic transition the heat conductivity $\lambda(B,T)$ has a pronounced drop, which often occur in AFM systems~\cite{metcalfe1972magnetothermal, zhao2012heat, dixon1980low}. The points of critical field at which the heat conductivity $\lambda(B,T)$ drops were determined from anomaly on the calculated heat capacity. The heat capacity $C_p(B,T)$ and the heat conductivity $\lambda(B,T)$ as functions of external field and temperature, as these were used in our calculations, can be seen in Fig.~\ref{hchc}.

\bibliography{main}

\begin{thebibliography}{41}%
\makeatletter
\providecommand \@ifxundefined [1]{%
 \@ifx{#1\undefined}
}%
\providecommand \@ifnum [1]{%
 \ifnum #1\expandafter \@firstoftwo
 \else \expandafter \@secondoftwo
 \fi
}%
\providecommand \@ifx [1]{%
 \ifx #1\expandafter \@firstoftwo
 \else \expandafter \@secondoftwo
 \fi
}%
\providecommand \natexlab [1]{#1}%
\providecommand \enquote  [1]{``#1''}%
\providecommand \bibnamefont  [1]{#1}%
\providecommand \bibfnamefont [1]{#1}%
\providecommand \citenamefont [1]{#1}%
\providecommand \href@noop [0]{\@secondoftwo}%
\providecommand \href [0]{\begingroup \@sanitize@url \@href}%
\providecommand \@href[1]{\@@startlink{#1}\@@href}%
\providecommand \@@href[1]{\endgroup#1\@@endlink}%
\providecommand \@sanitize@url [0]{\catcode `\\12\catcode `\$12\catcode
  `\&12\catcode `\#12\catcode `\^12\catcode `\_12\catcode `\%12\relax}%
\providecommand \@@startlink[1]{}%
\providecommand \@@endlink[0]{}%
\providecommand \url  [0]{\begingroup\@sanitize@url \@url }%
\providecommand \@url [1]{\endgroup\@href {#1}{\urlprefix }}%
\providecommand \urlprefix  [0]{URL }%
\providecommand \Eprint [0]{\href }%
\providecommand \doibase [0]{https://doi.org/}%
\providecommand \selectlanguage [0]{\@gobble}%
\providecommand \bibinfo  [0]{\@secondoftwo}%
\providecommand \bibfield  [0]{\@secondoftwo}%
\providecommand \translation [1]{[#1]}%
\providecommand \BibitemOpen [0]{}%
\providecommand \bibitemStop [0]{}%
\providecommand \bibitemNoStop [0]{.\EOS\space}%
\providecommand \EOS [0]{\spacefactor3000\relax}%
\providecommand \BibitemShut  [1]{\csname bibitem#1\endcsname}%
\let\auto@bib@innerbib\@empty
\bibitem [{\citenamefont {De~Jong}\ \emph {et~al.}(2011)\citenamefont
  {De~Jong}, \citenamefont {Kimel}, \citenamefont {Pisarev}, \citenamefont
  {Kirilyuk},\ and\ \citenamefont {Rasing}}]{de2011laser}%
  \BibitemOpen
  \bibfield  {author} {\bibinfo {author} {\bibfnamefont {J.}~\bibnamefont
  {De~Jong}}, \bibinfo {author} {\bibfnamefont {A.}~\bibnamefont {Kimel}},
  \bibinfo {author} {\bibfnamefont {R.}~\bibnamefont {Pisarev}}, \bibinfo
  {author} {\bibfnamefont {A.}~\bibnamefont {Kirilyuk}},\ and\ \bibinfo
  {author} {\bibfnamefont {T.}~\bibnamefont {Rasing}},\ }\bibfield  {title}
  {\bibinfo {title} {{Laser-induced ultrafast spin dynamics in ErFeO$_3$}},\
  }\href {https://doi.org/10.1103/PhysRevB.84.104421} {\bibfield  {journal}
  {\bibinfo  {journal} {Phys. Rev. B}\ }\textbf {\bibinfo {volume} {84}},\
  \bibinfo {pages} {104421} (\bibinfo {year} {2011})}\BibitemShut {NoStop}%
\bibitem [{\citenamefont {Neeraj}\ \emph {et~al.}(2021)\citenamefont {Neeraj},
  \citenamefont {Awari}, \citenamefont {Kovalev}, \citenamefont {Polley},
  \citenamefont {Zhou~Hagstr{\"o}m}, \citenamefont {Arekapudi}, \citenamefont
  {Semisalova}, \citenamefont {Lenz}, \citenamefont {Green}, \citenamefont
  {Deinert}, \citenamefont {Ilyakov}, \citenamefont {Chen}, \citenamefont
  {Bawatna}, \citenamefont {Scalera}, \citenamefont {d'Aquino}, \citenamefont
  {Serpico}, \citenamefont {Hellwig}, \citenamefont {Wegrowe}, \citenamefont
  {Gensch},\ and\ \citenamefont {Bonetti}}]{neeraj2021inertial}%
  \BibitemOpen
  \bibfield  {author} {\bibinfo {author} {\bibfnamefont {K.}~\bibnamefont
  {Neeraj}}, \bibinfo {author} {\bibfnamefont {N.}~\bibnamefont {Awari}},
  \bibinfo {author} {\bibfnamefont {S.}~\bibnamefont {Kovalev}}, \bibinfo
  {author} {\bibfnamefont {D.}~\bibnamefont {Polley}}, \bibinfo {author}
  {\bibfnamefont {N.}~\bibnamefont {Zhou~Hagstr{\"o}m}}, \bibinfo {author}
  {\bibfnamefont {S.~S. P.~K.}\ \bibnamefont {Arekapudi}}, \bibinfo {author}
  {\bibfnamefont {A.}~\bibnamefont {Semisalova}}, \bibinfo {author}
  {\bibfnamefont {K.}~\bibnamefont {Lenz}}, \bibinfo {author} {\bibfnamefont
  {B.}~\bibnamefont {Green}}, \bibinfo {author} {\bibfnamefont {J.-C.}\
  \bibnamefont {Deinert}}, \bibinfo {author} {\bibfnamefont {I.}~\bibnamefont
  {Ilyakov}}, \bibinfo {author} {\bibfnamefont {M.}~\bibnamefont {Chen}},
  \bibinfo {author} {\bibfnamefont {M.}~\bibnamefont {Bawatna}}, \bibinfo
  {author} {\bibfnamefont {V.}~\bibnamefont {Scalera}}, \bibinfo {author}
  {\bibfnamefont {M.}~\bibnamefont {d'Aquino}}, \bibinfo {author}
  {\bibfnamefont {C.}~\bibnamefont {Serpico}}, \bibinfo {author} {\bibfnamefont
  {O.}~\bibnamefont {Hellwig}}, \bibinfo {author} {\bibfnamefont {J.-E.}\
  \bibnamefont {Wegrowe}}, \bibinfo {author} {\bibfnamefont {M.}~\bibnamefont
  {Gensch}},\ and\ \bibinfo {author} {\bibfnamefont {S.}~\bibnamefont
  {Bonetti}},\ }\bibfield  {title} {\bibinfo {title} {Inertial spin dynamics in
  ferromagnets},\ }\href {https://doi.org/10.1038/s41567-020-01040-y}
  {\bibfield  {journal} {\bibinfo  {journal} {Nature Phys.}\ }\textbf {\bibinfo
  {volume} {17}},\ \bibinfo {pages} {245} (\bibinfo {year} {2021})}\BibitemShut
  {NoStop}%
\bibitem [{\citenamefont {de~la Torre}\ \emph {et~al.}(2021)\citenamefont
  {de~la Torre}, \citenamefont {Kennes}, \citenamefont {Claassen},
  \citenamefont {Gerber}, \citenamefont {McIver},\ and\ \citenamefont
  {Sentef}}]{de2021colloquium}%
  \BibitemOpen
  \bibfield  {author} {\bibinfo {author} {\bibfnamefont {A.}~\bibnamefont
  {de~la Torre}}, \bibinfo {author} {\bibfnamefont {D.~M.}\ \bibnamefont
  {Kennes}}, \bibinfo {author} {\bibfnamefont {M.}~\bibnamefont {Claassen}},
  \bibinfo {author} {\bibfnamefont {S.}~\bibnamefont {Gerber}}, \bibinfo
  {author} {\bibfnamefont {J.~W.}\ \bibnamefont {McIver}},\ and\ \bibinfo
  {author} {\bibfnamefont {M.~A.}\ \bibnamefont {Sentef}},\ }\bibfield  {title}
  {\bibinfo {title} {{Colloquium: Nonthermal pathways to ultrafast control in
  quantum materials}},\ }\href {https://doi.org/10.1103/RevModPhys.93.041002}
  {\bibfield  {journal} {\bibinfo  {journal} {Rev. Mod. Phys.}\ }\textbf
  {\bibinfo {volume} {93}},\ \bibinfo {pages} {041002} (\bibinfo {year}
  {2021})}\BibitemShut {NoStop}%
\bibitem [{\citenamefont {Gatteschi}\ and\ \citenamefont
  {Sessoli}(2003)}]{gatteschi2003quantum}%
  \BibitemOpen
  \bibfield  {author} {\bibinfo {author} {\bibfnamefont {D.}~\bibnamefont
  {Gatteschi}}\ and\ \bibinfo {author} {\bibfnamefont {R.}~\bibnamefont
  {Sessoli}},\ }\bibfield  {title} {\bibinfo {title} {Quantum tunneling of
  magnetization and related phenomena in molecular materials},\ }\href
  {https://doi.org/10.1002/anie.200390099} {\bibfield  {journal} {\bibinfo
  {journal} {Angew. Chem.}\ }\textbf {\bibinfo {volume} {42}},\ \bibinfo
  {pages} {268} (\bibinfo {year} {2003})}\BibitemShut {NoStop}%
\bibitem [{\citenamefont {Langley}\ \emph {et~al.}(2014)\citenamefont
  {Langley}, \citenamefont {Wielechowski}, \citenamefont {Vieru}, \citenamefont
  {Chilton}, \citenamefont {Moubaraki}, \citenamefont {Chibotaru},\ and\
  \citenamefont {Murray}}]{langley2014modulation}%
  \BibitemOpen
  \bibfield  {author} {\bibinfo {author} {\bibfnamefont {S.~K.}\ \bibnamefont
  {Langley}}, \bibinfo {author} {\bibfnamefont {D.~P.}\ \bibnamefont
  {Wielechowski}}, \bibinfo {author} {\bibfnamefont {V.}~\bibnamefont {Vieru}},
  \bibinfo {author} {\bibfnamefont {N.~F.}\ \bibnamefont {Chilton}}, \bibinfo
  {author} {\bibfnamefont {B.}~\bibnamefont {Moubaraki}}, \bibinfo {author}
  {\bibfnamefont {L.~F.}\ \bibnamefont {Chibotaru}},\ and\ \bibinfo {author}
  {\bibfnamefont {K.~S.}\ \bibnamefont {Murray}},\ }\bibfield  {title}
  {\bibinfo {title} {{Modulation of slow magnetic relaxation by tuning magnetic
  exchange in Cr$_2$Dy$_2$ single molecule magnets}},\ }\href
  {https://doi.org/10.1039/C4SC01239A} {\bibfield  {journal} {\bibinfo
  {journal} {Chem. Sci.}\ }\textbf {\bibinfo {volume} {5}},\ \bibinfo {pages}
  {3246} (\bibinfo {year} {2014})}\BibitemShut {NoStop}%
\bibitem [{\citenamefont {Guo}\ \emph {et~al.}(2011)\citenamefont {Guo},
  \citenamefont {Xu}, \citenamefont {Guo},\ and\ \citenamefont
  {Tang}}]{guo2011relaxation}%
  \BibitemOpen
  \bibfield  {author} {\bibinfo {author} {\bibfnamefont {Y.-N.}\ \bibnamefont
  {Guo}}, \bibinfo {author} {\bibfnamefont {G.-F.}\ \bibnamefont {Xu}},
  \bibinfo {author} {\bibfnamefont {Y.}~\bibnamefont {Guo}},\ and\ \bibinfo
  {author} {\bibfnamefont {J.}~\bibnamefont {Tang}},\ }\bibfield  {title}
  {\bibinfo {title} {Relaxation dynamics of dysprosium (iii) single molecule
  magnets},\ }\href {https://doi.org/10.1039/c1dt10474h} {\bibfield  {journal}
  {\bibinfo  {journal} {Dalton Transactions}\ }\textbf {\bibinfo {volume}
  {40}},\ \bibinfo {pages} {9953} (\bibinfo {year} {2011})}\BibitemShut
  {NoStop}%
\bibitem [{\citenamefont {Matsuhira}\ \emph {et~al.}(2001)\citenamefont
  {Matsuhira}, \citenamefont {Hinatsu},\ and\ \citenamefont
  {Sakakibara}}]{matsuhira2001novel}%
  \BibitemOpen
  \bibfield  {author} {\bibinfo {author} {\bibfnamefont {K.}~\bibnamefont
  {Matsuhira}}, \bibinfo {author} {\bibfnamefont {Y.}~\bibnamefont {Hinatsu}},\
  and\ \bibinfo {author} {\bibfnamefont {T.}~\bibnamefont {Sakakibara}},\
  }\bibfield  {title} {\bibinfo {title} {{Novel dynamical magnetic properties
  in the spin ice compound Dy$_2$Ti$_2$O$_7$}},\ }\href@noop {} {\bibfield
  {journal} {\bibinfo  {journal} {Journal of Physics: Condensed Matter}\
  }\textbf {\bibinfo {volume} {13}},\ \bibinfo {pages} {L737} (\bibinfo {year}
  {2001})}\BibitemShut {NoStop}%
\bibitem [{\citenamefont {Snyder}\ \emph {et~al.}(2004)\citenamefont {Snyder},
  \citenamefont {Ueland}, \citenamefont {Slusky}, \citenamefont {Karunadasa},
  \citenamefont {Cava},\ and\ \citenamefont {Schiffer}}]{snyder2004low}%
  \BibitemOpen
  \bibfield  {author} {\bibinfo {author} {\bibfnamefont {J.}~\bibnamefont
  {Snyder}}, \bibinfo {author} {\bibfnamefont {B.}~\bibnamefont {Ueland}},
  \bibinfo {author} {\bibfnamefont {J.}~\bibnamefont {Slusky}}, \bibinfo
  {author} {\bibfnamefont {H.}~\bibnamefont {Karunadasa}}, \bibinfo {author}
  {\bibfnamefont {R.}~\bibnamefont {Cava}},\ and\ \bibinfo {author}
  {\bibfnamefont {P.}~\bibnamefont {Schiffer}},\ }\bibfield  {title} {\bibinfo
  {title} {{Low-temperature spin freezing in the Dy$_2$Ti$_2$O$_7$ spin ice}},\
  }\href {https://doi.org/10.1103/PhysRevB.69.064414} {\bibfield  {journal}
  {\bibinfo  {journal} {Phys. Rev. B}\ }\textbf {\bibinfo {volume} {69}},\
  \bibinfo {pages} {064414} (\bibinfo {year} {2004})}\BibitemShut {NoStop}%
\bibitem [{\citenamefont {Matsuhira}\ \emph {et~al.}(2011)\citenamefont
  {Matsuhira}, \citenamefont {Paulsen}, \citenamefont {Lhotel}, \citenamefont
  {Sekine}, \citenamefont {Hiroi},\ and\ \citenamefont
  {Takagi}}]{matsuhira2011spin}%
  \BibitemOpen
  \bibfield  {author} {\bibinfo {author} {\bibfnamefont {K.}~\bibnamefont
  {Matsuhira}}, \bibinfo {author} {\bibfnamefont {C.}~\bibnamefont {Paulsen}},
  \bibinfo {author} {\bibfnamefont {E.}~\bibnamefont {Lhotel}}, \bibinfo
  {author} {\bibfnamefont {C.}~\bibnamefont {Sekine}}, \bibinfo {author}
  {\bibfnamefont {Z.}~\bibnamefont {Hiroi}},\ and\ \bibinfo {author}
  {\bibfnamefont {S.}~\bibnamefont {Takagi}},\ }\bibfield  {title} {\bibinfo
  {title} {{Spin dynamics at very low temperature in spin ice
  Dy$_2$Ti$_2$O$_7$}},\ }\href@noop {} {\bibfield  {journal} {\bibinfo
  {journal} {J. Phys. Soc. Jpn}\ }\textbf {\bibinfo {volume} {80}},\ \bibinfo
  {pages} {123711} (\bibinfo {year} {2011})}\BibitemShut {NoStop}%
\bibitem [{\citenamefont {Takatsu}\ \emph {et~al.}(2013)\citenamefont
  {Takatsu}, \citenamefont {Goto}, \citenamefont {Otsuka}, \citenamefont
  {Higashinaka}, \citenamefont {Matsubayashi}, \citenamefont {Uwatoko},\ and\
  \citenamefont {Kadowaki}}]{takatsu2013ac}%
  \BibitemOpen
  \bibfield  {author} {\bibinfo {author} {\bibfnamefont {H.}~\bibnamefont
  {Takatsu}}, \bibinfo {author} {\bibfnamefont {K.}~\bibnamefont {Goto}},
  \bibinfo {author} {\bibfnamefont {H.}~\bibnamefont {Otsuka}}, \bibinfo
  {author} {\bibfnamefont {R.}~\bibnamefont {Higashinaka}}, \bibinfo {author}
  {\bibfnamefont {K.}~\bibnamefont {Matsubayashi}}, \bibinfo {author}
  {\bibfnamefont {Y.}~\bibnamefont {Uwatoko}},\ and\ \bibinfo {author}
  {\bibfnamefont {H.}~\bibnamefont {Kadowaki}},\ }\bibfield  {title} {\bibinfo
  {title} {{AC susceptibility of the dipolar spin ice Dy$_2$Ti$_2$O$_7$:
  Experiments and Monte Carlo simulations}},\ }\href
  {https://doi.org/10.7566/JPSJ.82.104710} {\bibfield  {journal} {\bibinfo
  {journal} {J. Phys. Soc. Jpn}\ }\textbf {\bibinfo {volume} {82}},\ \bibinfo
  {pages} {104710} (\bibinfo {year} {2013})}\BibitemShut {NoStop}%
\bibitem [{\citenamefont {Gardner}\ \emph {et~al.}(2011)\citenamefont
  {Gardner}, \citenamefont {Ehlers}, \citenamefont {Fouquet}, \citenamefont
  {Farago},\ and\ \citenamefont {Stewart}}]{gardner2011slow}%
  \BibitemOpen
  \bibfield  {author} {\bibinfo {author} {\bibfnamefont {J.}~\bibnamefont
  {Gardner}}, \bibinfo {author} {\bibfnamefont {G.}~\bibnamefont {Ehlers}},
  \bibinfo {author} {\bibfnamefont {P.}~\bibnamefont {Fouquet}}, \bibinfo
  {author} {\bibfnamefont {B.}~\bibnamefont {Farago}},\ and\ \bibinfo {author}
  {\bibfnamefont {J.~R.}\ \bibnamefont {Stewart}},\ }\bibfield  {title}
  {\bibinfo {title} {{Slow and static spin correlations in
  Dy$_{2+x}$Ti$_{2-x}$O$_{7-\delta}$}},\ }\href
  {https://iopscience.iop.org/article/10.1088/0953-8984/23/16/164220/meta}
  {\bibfield  {journal} {\bibinfo  {journal} {J. Condens. Matter Phys.}\
  }\textbf {\bibinfo {volume} {23}},\ \bibinfo {pages} {164220} (\bibinfo
  {year} {2011})}\BibitemShut {NoStop}%
\bibitem [{\citenamefont {Ehlers}\ \emph {et~al.}(2002)\citenamefont {Ehlers},
  \citenamefont {Cornelius}, \citenamefont {Orendac}, \citenamefont
  {Kajnakova}, \citenamefont {Fennell}, \citenamefont {Bramwell},\ and\
  \citenamefont {Gardner}}]{ehlers2002dynamical}%
  \BibitemOpen
  \bibfield  {author} {\bibinfo {author} {\bibfnamefont {G.}~\bibnamefont
  {Ehlers}}, \bibinfo {author} {\bibfnamefont {A.}~\bibnamefont {Cornelius}},
  \bibinfo {author} {\bibfnamefont {M.}~\bibnamefont {Orendac}}, \bibinfo
  {author} {\bibfnamefont {M.}~\bibnamefont {Kajnakova}}, \bibinfo {author}
  {\bibfnamefont {T.}~\bibnamefont {Fennell}}, \bibinfo {author} {\bibfnamefont
  {S.}~\bibnamefont {Bramwell}},\ and\ \bibinfo {author} {\bibfnamefont
  {J.}~\bibnamefont {Gardner}},\ }\bibfield  {title} {\bibinfo {title}
  {{Dynamical crossover in 'hot' spin ice}},\ }\href
  {https://iopscience.iop.org/article/10.1088/0953-8984/15/2/102/meta}
  {\bibfield  {journal} {\bibinfo  {journal} {J. Condens. Matter Phys.}\
  }\textbf {\bibinfo {volume} {15}},\ \bibinfo {pages} {L9} (\bibinfo {year}
  {2002})}\BibitemShut {NoStop}%
\bibitem [{\citenamefont {Hardy}\ \emph {et~al.}(2004)\citenamefont {Hardy},
  \citenamefont {Flahaut}, \citenamefont {Lees},\ and\ \citenamefont
  {Petrenko}}]{hardy2004magnetic}%
  \BibitemOpen
  \bibfield  {author} {\bibinfo {author} {\bibfnamefont {V.}~\bibnamefont
  {Hardy}}, \bibinfo {author} {\bibfnamefont {D.}~\bibnamefont {Flahaut}},
  \bibinfo {author} {\bibfnamefont {M.}~\bibnamefont {Lees}},\ and\ \bibinfo
  {author} {\bibfnamefont {O.}~\bibnamefont {Petrenko}},\ }\bibfield  {title}
  {\bibinfo {title} {{Magnetic quantum tunneling in Ca$_3$Co$_2$O$_6$ studied
  by AC susceptibility: Temperature and magnetic-field dependence of the
  spin-relaxation time}},\ }\href@noop {} {\bibfield  {journal} {\bibinfo
  {journal} {Phys. Rev. B}\ }\textbf {\bibinfo {volume} {70}},\ \bibinfo
  {pages} {214439} (\bibinfo {year} {2004})}\BibitemShut {NoStop}%
\bibitem [{\citenamefont {Hegde}\ \emph {et~al.}(2020)\citenamefont {Hegde},
  \citenamefont {Levati{\'c}}, \citenamefont {Magrez}, \citenamefont
  {R{\o}nnow},\ and\ \citenamefont {{\v{Z}}ivkovi{\'c}}}]{hegde2020magnetic}%
  \BibitemOpen
  \bibfield  {author} {\bibinfo {author} {\bibfnamefont {N.~G.}\ \bibnamefont
  {Hegde}}, \bibinfo {author} {\bibfnamefont {I.}~\bibnamefont {Levati{\'c}}},
  \bibinfo {author} {\bibfnamefont {A.}~\bibnamefont {Magrez}}, \bibinfo
  {author} {\bibfnamefont {H.~M.}\ \bibnamefont {R{\o}nnow}},\ and\ \bibinfo
  {author} {\bibfnamefont {I.}~\bibnamefont {{\v{Z}}ivkovi{\'c}}},\ }\bibfield
  {title} {\bibinfo {title} {{Magnetic dynamics across the in-field transition
  in Ca$_3$Co$_2$O$_6$}},\ }\href {https://doi.org/10.1103/PhysRevB.102.104418}
  {\bibfield  {journal} {\bibinfo  {journal} {Phys. Rev. B}\ }\textbf {\bibinfo
  {volume} {102}},\ \bibinfo {pages} {104418} (\bibinfo {year}
  {2020})}\BibitemShut {NoStop}%
\bibitem [{\citenamefont {Bitko}\ \emph {et~al.}(1996)\citenamefont {Bitko},
  \citenamefont {Rosenbaum},\ and\ \citenamefont {Aeppli}}]{bitko1996quantum}%
  \BibitemOpen
  \bibfield  {author} {\bibinfo {author} {\bibfnamefont {D.}~\bibnamefont
  {Bitko}}, \bibinfo {author} {\bibfnamefont {T.}~\bibnamefont {Rosenbaum}},\
  and\ \bibinfo {author} {\bibfnamefont {G.}~\bibnamefont {Aeppli}},\
  }\bibfield  {title} {\bibinfo {title} {Quantum critical behavior for a model
  magnet},\ }\href {https://doi.org/10.1103/PhysRevLett.77.940} {\bibfield
  {journal} {\bibinfo  {journal} {Phys. Rev. Lett.}\ }\textbf {\bibinfo
  {volume} {77}},\ \bibinfo {pages} {940} (\bibinfo {year} {1996})}\BibitemShut
  {NoStop}%
\bibitem [{\citenamefont {Giraud}\ \emph {et~al.}(2001)\citenamefont {Giraud},
  \citenamefont {Wernsdorfer}, \citenamefont {Tkachuk}, \citenamefont
  {Mailly},\ and\ \citenamefont {Barbara}}]{giraud2001nuclear}%
  \BibitemOpen
  \bibfield  {author} {\bibinfo {author} {\bibfnamefont {R.}~\bibnamefont
  {Giraud}}, \bibinfo {author} {\bibfnamefont {W.}~\bibnamefont {Wernsdorfer}},
  \bibinfo {author} {\bibfnamefont {A.}~\bibnamefont {Tkachuk}}, \bibinfo
  {author} {\bibfnamefont {D.}~\bibnamefont {Mailly}},\ and\ \bibinfo {author}
  {\bibfnamefont {B.}~\bibnamefont {Barbara}},\ }\bibfield  {title} {\bibinfo
  {title} {{Nuclear spin driven quantum relaxation in
  LiY$_{0.998}$Ho$_{0.002}$F$_4$}},\ }\href
  {https://doi.org/10.1103/PhysRevLett.87.057203} {\bibfield  {journal}
  {\bibinfo  {journal} {Phys. Rev. Lett.}\ }\textbf {\bibinfo {volume} {87}},\
  \bibinfo {pages} {057203} (\bibinfo {year} {2001})}\BibitemShut {NoStop}%
\bibitem [{\citenamefont {Bertaina}\ \emph {et~al.}(2006)\citenamefont
  {Bertaina}, \citenamefont {Barbara}, \citenamefont {Giraud}, \citenamefont
  {Malkin}, \citenamefont {Vanuynin}, \citenamefont {Pominov}, \citenamefont
  {Stolov},\ and\ \citenamefont {Tkachuk}}]{bertaina2006cross}%
  \BibitemOpen
  \bibfield  {author} {\bibinfo {author} {\bibfnamefont {S.}~\bibnamefont
  {Bertaina}}, \bibinfo {author} {\bibfnamefont {B.}~\bibnamefont {Barbara}},
  \bibinfo {author} {\bibfnamefont {R.}~\bibnamefont {Giraud}}, \bibinfo
  {author} {\bibfnamefont {B.}~\bibnamefont {Malkin}}, \bibinfo {author}
  {\bibfnamefont {M.}~\bibnamefont {Vanuynin}}, \bibinfo {author}
  {\bibfnamefont {A.}~\bibnamefont {Pominov}}, \bibinfo {author} {\bibfnamefont
  {A.}~\bibnamefont {Stolov}},\ and\ \bibinfo {author} {\bibfnamefont
  {A.}~\bibnamefont {Tkachuk}},\ }\bibfield  {title} {\bibinfo {title}
  {{Cross-relaxation and phonon bottleneck effects on magnetization dynamics in
  LiYF$_4$:Ho$^{3+}$}},\ }\href {https://doi.org/10.1103/PhysRevB.74.184421}
  {\bibfield  {journal} {\bibinfo  {journal} {Phys. Rev. B}\ }\textbf {\bibinfo
  {volume} {74}},\ \bibinfo {pages} {184421} (\bibinfo {year}
  {2006})}\BibitemShut {NoStop}%
\bibitem [{\citenamefont {Johnson}\ \emph {et~al.}(2012)\citenamefont
  {Johnson}, \citenamefont {Malkin}, \citenamefont {Lord}, \citenamefont
  {Giblin}, \citenamefont {Amato}, \citenamefont {Baines}, \citenamefont
  {Lascialfari}, \citenamefont {Barbara},\ and\ \citenamefont
  {Graf}}]{johnson2012evolution}%
  \BibitemOpen
  \bibfield  {author} {\bibinfo {author} {\bibfnamefont {R.}~\bibnamefont
  {Johnson}}, \bibinfo {author} {\bibfnamefont {B.}~\bibnamefont {Malkin}},
  \bibinfo {author} {\bibfnamefont {J.}~\bibnamefont {Lord}}, \bibinfo {author}
  {\bibfnamefont {S.}~\bibnamefont {Giblin}}, \bibinfo {author} {\bibfnamefont
  {A.}~\bibnamefont {Amato}}, \bibinfo {author} {\bibfnamefont
  {C.}~\bibnamefont {Baines}}, \bibinfo {author} {\bibfnamefont
  {A.}~\bibnamefont {Lascialfari}}, \bibinfo {author} {\bibfnamefont
  {B.}~\bibnamefont {Barbara}},\ and\ \bibinfo {author} {\bibfnamefont {M.~J.}\
  \bibnamefont {Graf}},\ }\bibfield  {title} {\bibinfo {title} {{Evolution of
  spin relaxation processes in LiY$_{1-x}$Ho$_x$F$_4$ studied via
  ac-susceptibility and muon spin relaxation}},\ }\href
  {https://doi.org/10.1103/PhysRevB.86.014427} {\bibfield  {journal} {\bibinfo
  {journal} {Phys. Rev. B}\ }\textbf {\bibinfo {volume} {86}},\ \bibinfo
  {pages} {014427} (\bibinfo {year} {2012})}\BibitemShut {NoStop}%
\bibitem [{\citenamefont {Wu}\ \emph {et~al.}(2017)\citenamefont {Wu},
  \citenamefont {Nikitin}, \citenamefont {Frontzek}, \citenamefont
  {Kolesnikov}, \citenamefont {Ehlers}, \citenamefont {Lumsden}, \citenamefont
  {Shaykhutdinov}, \citenamefont {Guo}, \citenamefont {Savici}, \citenamefont
  {Gai} \emph {et~al.}}]{wu2017magnetic}%
  \BibitemOpen
  \bibfield  {author} {\bibinfo {author} {\bibfnamefont {L.}~\bibnamefont
  {Wu}}, \bibinfo {author} {\bibfnamefont {S.~E.}\ \bibnamefont {Nikitin}},
  \bibinfo {author} {\bibfnamefont {M.}~\bibnamefont {Frontzek}}, \bibinfo
  {author} {\bibfnamefont {A.~I.}\ \bibnamefont {Kolesnikov}}, \bibinfo
  {author} {\bibfnamefont {G.}~\bibnamefont {Ehlers}}, \bibinfo {author}
  {\bibfnamefont {M.~D.}\ \bibnamefont {Lumsden}}, \bibinfo {author}
  {\bibfnamefont {K.~A.}\ \bibnamefont {Shaykhutdinov}}, \bibinfo {author}
  {\bibfnamefont {E.-J.}\ \bibnamefont {Guo}}, \bibinfo {author} {\bibfnamefont
  {A.~T.}\ \bibnamefont {Savici}}, \bibinfo {author} {\bibfnamefont
  {Z.}~\bibnamefont {Gai}}, \emph {et~al.},\ }\bibfield  {title} {\bibinfo
  {title} {{Magnetic ground state of the Ising-like antiferromagnet
  DyScO$_3$}},\ }\href {https://doi.org/10.1103/PhysRevB.96.144407} {\bibfield
  {journal} {\bibinfo  {journal} {Phys. Rev. B}\ }\textbf {\bibinfo {volume}
  {96}},\ \bibinfo {pages} {144407} (\bibinfo {year} {2017})}\BibitemShut
  {NoStop}%
\bibitem [{\citenamefont {Ke}\ \emph {et~al.}(2009)\citenamefont {Ke},
  \citenamefont {Adamo}, \citenamefont {Schlom}, \citenamefont {Bernhagen},
  \citenamefont {Uecker},\ and\ \citenamefont {Schiffer}}]{ke2009low}%
  \BibitemOpen
  \bibfield  {author} {\bibinfo {author} {\bibfnamefont {X.}~\bibnamefont
  {Ke}}, \bibinfo {author} {\bibfnamefont {C.}~\bibnamefont {Adamo}}, \bibinfo
  {author} {\bibfnamefont {D.~G.}\ \bibnamefont {Schlom}}, \bibinfo {author}
  {\bibfnamefont {M.}~\bibnamefont {Bernhagen}}, \bibinfo {author}
  {\bibfnamefont {R.}~\bibnamefont {Uecker}},\ and\ \bibinfo {author}
  {\bibfnamefont {P.}~\bibnamefont {Schiffer}},\ }\bibfield  {title} {\bibinfo
  {title} {{Low temperature magnetism in the perovskite substrate DyScO$_3$}},\
  }\href {https://doi.org/10.1063/1.3117190} {\bibfield  {journal} {\bibinfo
  {journal} {Appl. Phys. Lett.}\ }\textbf {\bibinfo {volume} {94}},\ \bibinfo
  {pages} {152503} (\bibinfo {year} {2009})}\BibitemShut {NoStop}%
\bibitem [{\citenamefont {Wu}\ \emph {et~al.}(2019{\natexlab{a}})\citenamefont
  {Wu}, \citenamefont {Qin}, \citenamefont {Ma}, \citenamefont {Li},
  \citenamefont {Wei},\ and\ \citenamefont {Zi}}]{wu2019large}%
  \BibitemOpen
  \bibfield  {author} {\bibinfo {author} {\bibfnamefont {Y.-D.}\ \bibnamefont
  {Wu}}, \bibinfo {author} {\bibfnamefont {Y.-L.}\ \bibnamefont {Qin}},
  \bibinfo {author} {\bibfnamefont {X.-H.}\ \bibnamefont {Ma}}, \bibinfo
  {author} {\bibfnamefont {R.-W.}\ \bibnamefont {Li}}, \bibinfo {author}
  {\bibfnamefont {Y.-Y.}\ \bibnamefont {Wei}},\ and\ \bibinfo {author}
  {\bibfnamefont {Z.-F.}\ \bibnamefont {Zi}},\ }\bibfield  {title} {\bibinfo
  {title} {{Large rotating magnetocaloric effect at low magnetic fields in the
  Ising-like antiferromagnet DyScO$_3$ single crystal}},\ }\href
  {https://doi.org/10.1016/j.jallcom.2018.11.047} {\bibfield  {journal}
  {\bibinfo  {journal} {J. Alloys Compd.}\ }\textbf {\bibinfo {volume} {777}},\
  \bibinfo {pages} {673} (\bibinfo {year} {2019}{\natexlab{a}})}\BibitemShut
  {NoStop}%
\bibitem [{\citenamefont {Bluschke}\ \emph {et~al.}(2017)\citenamefont
  {Bluschke}, \citenamefont {Frano}, \citenamefont {Schierle}, \citenamefont
  {Minola}, \citenamefont {Hepting}, \citenamefont {Christiani}, \citenamefont
  {Logvenov}, \citenamefont {Weschke}, \citenamefont {Benckiser},\ and\
  \citenamefont {Keimer}}]{bluschke2017transfer}%
  \BibitemOpen
  \bibfield  {author} {\bibinfo {author} {\bibfnamefont {M.}~\bibnamefont
  {Bluschke}}, \bibinfo {author} {\bibfnamefont {A.}~\bibnamefont {Frano}},
  \bibinfo {author} {\bibfnamefont {E.}~\bibnamefont {Schierle}}, \bibinfo
  {author} {\bibfnamefont {M.}~\bibnamefont {Minola}}, \bibinfo {author}
  {\bibfnamefont {M.}~\bibnamefont {Hepting}}, \bibinfo {author} {\bibfnamefont
  {G.}~\bibnamefont {Christiani}}, \bibinfo {author} {\bibfnamefont
  {G.}~\bibnamefont {Logvenov}}, \bibinfo {author} {\bibfnamefont
  {E.}~\bibnamefont {Weschke}}, \bibinfo {author} {\bibfnamefont
  {E.}~\bibnamefont {Benckiser}},\ and\ \bibinfo {author} {\bibfnamefont
  {B.}~\bibnamefont {Keimer}},\ }\bibfield  {title} {\bibinfo {title}
  {{Transfer of magnetic order and anisotropy through epitaxial integration of
  3d and 4f spin systems}},\ }\href
  {https://doi.org/10.1103/PhysRevLett.118.207203} {\bibfield  {journal}
  {\bibinfo  {journal} {Phys. Rev. Lett.}\ }\textbf {\bibinfo {volume} {118}},\
  \bibinfo {pages} {207203} (\bibinfo {year} {2017})}\BibitemShut {NoStop}%
\bibitem [{\citenamefont {Kroder}\ \emph {et~al.}(2019)\citenamefont {Kroder},
  \citenamefont {Manna}, \citenamefont {Kriegner}, \citenamefont {Sukhanov},
  \citenamefont {Liu}, \citenamefont {Borrmann}, \citenamefont {Hoser},
  \citenamefont {Gooth}, \citenamefont {Schnelle}, \citenamefont {Inosov} \emph
  {et~al.}}]{kroder2019spin}%
  \BibitemOpen
  \bibfield  {author} {\bibinfo {author} {\bibfnamefont {J.}~\bibnamefont
  {Kroder}}, \bibinfo {author} {\bibfnamefont {K.}~\bibnamefont {Manna}},
  \bibinfo {author} {\bibfnamefont {D.}~\bibnamefont {Kriegner}}, \bibinfo
  {author} {\bibfnamefont {A.}~\bibnamefont {Sukhanov}}, \bibinfo {author}
  {\bibfnamefont {E.}~\bibnamefont {Liu}}, \bibinfo {author} {\bibfnamefont
  {H.}~\bibnamefont {Borrmann}}, \bibinfo {author} {\bibfnamefont
  {A.}~\bibnamefont {Hoser}}, \bibinfo {author} {\bibfnamefont
  {J.}~\bibnamefont {Gooth}}, \bibinfo {author} {\bibfnamefont
  {W.}~\bibnamefont {Schnelle}}, \bibinfo {author} {\bibfnamefont {D.~S.}\
  \bibnamefont {Inosov}}, \emph {et~al.},\ }\bibfield  {title} {\bibinfo
  {title} {Spin glass behavior in the disordered half-heusler compound
  irmnga},\ }\href {https://doi.org/10.1103/PhysRevB.99.174410} {\bibfield
  {journal} {\bibinfo  {journal} {Phys. Rev. B}\ }\textbf {\bibinfo {volume}
  {99}},\ \bibinfo {pages} {174410} (\bibinfo {year} {2019})}\BibitemShut
  {NoStop}%
\bibitem [{\citenamefont {Anand}\ \emph {et~al.}(2012)\citenamefont {Anand},
  \citenamefont {Adroja},\ and\ \citenamefont
  {Hillier}}]{anand2012ferromagnetic}%
  \BibitemOpen
  \bibfield  {author} {\bibinfo {author} {\bibfnamefont {V.}~\bibnamefont
  {Anand}}, \bibinfo {author} {\bibfnamefont {D.}~\bibnamefont {Adroja}},\ and\
  \bibinfo {author} {\bibfnamefont {A.}~\bibnamefont {Hillier}},\ }\bibfield
  {title} {\bibinfo {title} {{Ferromagnetic cluster spin-glass behavior in
  PrRhSn$_3$}},\ }\href {https://doi.org/10.1103/PhysRevB.85.014418} {\bibfield
   {journal} {\bibinfo  {journal} {Phys. Rev. B}\ }\textbf {\bibinfo {volume}
  {85}},\ \bibinfo {pages} {014418} (\bibinfo {year} {2012})}\BibitemShut
  {NoStop}%
\bibitem [{\citenamefont {Quilliam}\ \emph {et~al.}(2008)\citenamefont
  {Quilliam}, \citenamefont {Meng}, \citenamefont {Mugford},\ and\
  \citenamefont {Kycia}}]{quilliam2008evidence}%
  \BibitemOpen
  \bibfield  {author} {\bibinfo {author} {\bibfnamefont {J.}~\bibnamefont
  {Quilliam}}, \bibinfo {author} {\bibfnamefont {S.}~\bibnamefont {Meng}},
  \bibinfo {author} {\bibfnamefont {C.}~\bibnamefont {Mugford}},\ and\ \bibinfo
  {author} {\bibfnamefont {J.}~\bibnamefont {Kycia}},\ }\bibfield  {title}
  {\bibinfo {title} {{Evidence of spin glass dynamics in dilute
  LiHo$_x$Y$_{1-x}$F$_4$}},\ }\href
  {https://doi.org/10.1103/PhysRevLett.101.187204} {\bibfield  {journal}
  {\bibinfo  {journal} {Phys. Rev. Lett.}\ }\textbf {\bibinfo {volume} {101}},\
  \bibinfo {pages} {187204} (\bibinfo {year} {2008})}\BibitemShut {NoStop}%
\bibitem [{\citenamefont {Havriliak}\ and\ \citenamefont
  {Negami}(1967)}]{havriliak1967complex}%
  \BibitemOpen
  \bibfield  {author} {\bibinfo {author} {\bibfnamefont {S.}~\bibnamefont
  {Havriliak}}\ and\ \bibinfo {author} {\bibfnamefont {S.}~\bibnamefont
  {Negami}},\ }\bibfield  {title} {\bibinfo {title} {A complex plane
  representation of dielectric and mechanical relaxation processes in some
  polymers},\ }\href {https://doi.org/10.1016/0032-3861(67)90021-3} {\bibfield
  {journal} {\bibinfo  {journal} {Polymer}\ }\textbf {\bibinfo {volume} {8}},\
  \bibinfo {pages} {161} (\bibinfo {year} {1967})}\BibitemShut {NoStop}%
\bibitem [{\citenamefont {Topping}\ and\ \citenamefont
  {Blundell}(2018)}]{topping2018ac}%
  \BibitemOpen
  \bibfield  {author} {\bibinfo {author} {\bibfnamefont {C.}~\bibnamefont
  {Topping}}\ and\ \bibinfo {author} {\bibfnamefont {S.}~\bibnamefont
  {Blundell}},\ }\bibfield  {title} {\bibinfo {title} {Ac susceptibility as a
  probe of low-frequency magnetic dynamics},\ }\href@noop {} {\bibfield
  {journal} {\bibinfo  {journal} {J. Phys. Condens. Matter}\ }\textbf {\bibinfo
  {volume} {31}},\ \bibinfo {pages} {013001} (\bibinfo {year}
  {2018})}\BibitemShut {NoStop}%
\bibitem [{\citenamefont {Miskinis}(2009)}]{miskinis2009havriliak}%
  \BibitemOpen
  \bibfield  {author} {\bibinfo {author} {\bibfnamefont {P.}~\bibnamefont
  {Miskinis}},\ }\bibfield  {title} {\bibinfo {title} {The havriliak--negami
  susceptibility as a nonlinear and nonlocal process},\ }\href
  {https://doi.org/10.1088/0031-8949/2009/T136/014019} {\bibfield  {journal}
  {\bibinfo  {journal} {Phys. Scr.}\ }\textbf {\bibinfo {volume} {2009}},\
  \bibinfo {pages} {014019} (\bibinfo {year} {2009})}\BibitemShut {NoStop}%
\bibitem [{\citenamefont {Wu}\ \emph {et~al.}(2019{\natexlab{b}})\citenamefont
  {Wu}, \citenamefont {Nikitin}, \citenamefont {Brando}, \citenamefont
  {Vasylechko}, \citenamefont {Ehlers}, \citenamefont {Frontzek}, \citenamefont
  {Savici}, \citenamefont {Sala}, \citenamefont {Christianson}, \citenamefont
  {Lumsden},\ and\ \citenamefont {Podlesnyak}}]{wu2019antiferromagnetic}%
  \BibitemOpen
  \bibfield  {author} {\bibinfo {author} {\bibfnamefont {L.}~\bibnamefont
  {Wu}}, \bibinfo {author} {\bibfnamefont {S.}~\bibnamefont {Nikitin}},
  \bibinfo {author} {\bibfnamefont {M.}~\bibnamefont {Brando}}, \bibinfo
  {author} {\bibfnamefont {L.}~\bibnamefont {Vasylechko}}, \bibinfo {author}
  {\bibfnamefont {G.}~\bibnamefont {Ehlers}}, \bibinfo {author} {\bibfnamefont
  {M.}~\bibnamefont {Frontzek}}, \bibinfo {author} {\bibfnamefont {A.~T.}\
  \bibnamefont {Savici}}, \bibinfo {author} {\bibfnamefont {G.}~\bibnamefont
  {Sala}}, \bibinfo {author} {\bibfnamefont {A.~D.}\ \bibnamefont
  {Christianson}}, \bibinfo {author} {\bibfnamefont {M.~D.}\ \bibnamefont
  {Lumsden}},\ and\ \bibinfo {author} {\bibfnamefont {A.}~\bibnamefont
  {Podlesnyak}},\ }\bibfield  {title} {\bibinfo {title} {{Antiferromagnetic
  ordering and dipolar interactions of YbAlO$_3$}},\ }\href
  {https://doi.org/10.1103/PhysRevB.99.195117} {\bibfield  {journal} {\bibinfo
  {journal} {Phys. Rev. B}\ }\textbf {\bibinfo {volume} {99}},\ \bibinfo
  {pages} {195117} (\bibinfo {year} {2019}{\natexlab{b}})}\BibitemShut
  {NoStop}%
\bibitem [{\citenamefont {Beichl}\ and\ \citenamefont
  {Sullivan}(2000)}]{beichl2000metropolis}%
  \BibitemOpen
  \bibfield  {author} {\bibinfo {author} {\bibfnamefont {I.}~\bibnamefont
  {Beichl}}\ and\ \bibinfo {author} {\bibfnamefont {F.}~\bibnamefont
  {Sullivan}},\ }\bibfield  {title} {\bibinfo {title} {The metropolis
  algorithm},\ }\href {https://doi.org/10.1109/5992.814660} {\bibfield
  {journal} {\bibinfo  {journal} {Comput. Sci. Eng .}\ }\textbf {\bibinfo
  {volume} {2}},\ \bibinfo {pages} {65} (\bibinfo {year} {2000})}\BibitemShut
  {NoStop}%
\bibitem [{\citenamefont {Numazawa}\ \emph {et~al.}(1998)\citenamefont
  {Numazawa}, \citenamefont {Kimura}, \citenamefont {Shimamura},\ and\
  \citenamefont {Fukuda}}]{numazawa1998thermal}%
  \BibitemOpen
  \bibfield  {author} {\bibinfo {author} {\bibfnamefont {T.}~\bibnamefont
  {Numazawa}}, \bibinfo {author} {\bibfnamefont {H.}~\bibnamefont {Kimura}},
  \bibinfo {author} {\bibfnamefont {K.}~\bibnamefont {Shimamura}},\ and\
  \bibinfo {author} {\bibfnamefont {T.}~\bibnamefont {Fukuda}},\ }\bibfield
  {title} {\bibinfo {title} {{Thermal conductivity of $R$AlO$_3$ ($R=$ Dy, Er
  and Ho) in liquid helium temperatures}},\ }\href
  {https://doi.org/10.1023/A:1004326820468} {\bibfield  {journal} {\bibinfo
  {journal} {J. Mater. Sci.}\ }\textbf {\bibinfo {volume} {33}},\ \bibinfo
  {pages} {827} (\bibinfo {year} {1998})}\BibitemShut {NoStop}%
\bibitem [{\citenamefont {Metcalfe}\ and\ \citenamefont
  {Rosenberg}(1972)}]{metcalfe1972magnetothermal}%
  \BibitemOpen
  \bibfield  {author} {\bibinfo {author} {\bibfnamefont {M.~J.}\ \bibnamefont
  {Metcalfe}}\ and\ \bibinfo {author} {\bibfnamefont {H.~M.}\ \bibnamefont
  {Rosenberg}},\ }\bibfield  {title} {\bibinfo {title} {{The magnetothermal
  resistivity of antiferromagnetic crystals at low temperatures. I. DyPO$_4$, a
  nearly ideal Ising system}},\ }\href
  {https://iopscience.iop.org/article/10.1088/0022-3719/5/4/011/meta?casa_token=hpTYYXY4EgYAAAAA:5uSy-CDKOzX9h7qqoxoRbdD2p9PROAwOBcOWGNrnt5WLSEl5yaJXdUH6Wurx8gf93zl2ojOT2A}
  {\bibfield  {journal} {\bibinfo  {journal} {J. Phys. C Solid State Phys.}\
  }\textbf {\bibinfo {volume} {5}},\ \bibinfo {pages} {450} (\bibinfo {year}
  {1972})}\BibitemShut {NoStop}%
\bibitem [{\citenamefont {Zhao}\ \emph {et~al.}(2012)\citenamefont {Zhao},
  \citenamefont {Liu}, \citenamefont {He}, \citenamefont {Wang}, \citenamefont
  {Fan}, \citenamefont {Ke}, \citenamefont {Li}, \citenamefont {Chen},
  \citenamefont {Zhao},\ and\ \citenamefont {Sun}}]{zhao2012heat}%
  \BibitemOpen
  \bibfield  {author} {\bibinfo {author} {\bibfnamefont {Z.~Y.}\ \bibnamefont
  {Zhao}}, \bibinfo {author} {\bibfnamefont {X.~G.}\ \bibnamefont {Liu}},
  \bibinfo {author} {\bibfnamefont {Z.~Z.}\ \bibnamefont {He}}, \bibinfo
  {author} {\bibfnamefont {X.~M.}\ \bibnamefont {Wang}}, \bibinfo {author}
  {\bibfnamefont {C.}~\bibnamefont {Fan}}, \bibinfo {author} {\bibfnamefont
  {W.~P.}\ \bibnamefont {Ke}}, \bibinfo {author} {\bibfnamefont {Q.~J.}\
  \bibnamefont {Li}}, \bibinfo {author} {\bibfnamefont {L.~M.}\ \bibnamefont
  {Chen}}, \bibinfo {author} {\bibfnamefont {X.}~\bibnamefont {Zhao}},\ and\
  \bibinfo {author} {\bibfnamefont {X.~F.}\ \bibnamefont {Sun}},\ }\bibfield
  {title} {\bibinfo {title} {Heat transport of the quasi-one-dimensional
  ising-like antiferromagnet baco 2 v 2 o 8 in longitudinal and transverse
  fields},\ }\href {https://doi.org/10.1103/PhysRevB.85.134412} {\bibfield
  {journal} {\bibinfo  {journal} {Phys. Rev. B}\ }\textbf {\bibinfo {volume}
  {85}},\ \bibinfo {pages} {134412} (\bibinfo {year} {2012})}\BibitemShut
  {NoStop}%
\bibitem [{\citenamefont {Dixon}\ \emph {et~al.}(1980)\citenamefont {Dixon},
  \citenamefont {Benedict},\ and\ \citenamefont {Rives}}]{dixon1980low}%
  \BibitemOpen
  \bibfield  {author} {\bibinfo {author} {\bibfnamefont {G.~S.}\ \bibnamefont
  {Dixon}}, \bibinfo {author} {\bibfnamefont {V.}~\bibnamefont {Benedict}},\
  and\ \bibinfo {author} {\bibfnamefont {J.~E.}\ \bibnamefont {Rives}},\
  }\bibfield  {title} {\bibinfo {title} {{Low-temperature thermal conductivity
  of antiferromagnetic MnCl$_2\cdot$ 4H$_2$O}},\ }\href
  {https://doi.org/10.1103/PhysRevB.21.2865} {\bibfield  {journal} {\bibinfo
  {journal} {Phys. Rev. B}\ }\textbf {\bibinfo {volume} {21}},\ \bibinfo
  {pages} {2865} (\bibinfo {year} {1980})}\BibitemShut {NoStop}%
\bibitem [{\citenamefont {Fisher}\ and\ \citenamefont
  {Selke}(1980)}]{fisher1980infinitely}%
  \BibitemOpen
  \bibfield  {author} {\bibinfo {author} {\bibfnamefont {M.~E.}\ \bibnamefont
  {Fisher}}\ and\ \bibinfo {author} {\bibfnamefont {W.}~\bibnamefont {Selke}},\
  }\bibfield  {title} {\bibinfo {title} {Infinitely many commensurate phases in
  a simple ising model},\ }\href {https://doi.org/10.1103/PhysRevLett.44.1502}
  {\bibfield  {journal} {\bibinfo  {journal} {Phys. Rev. Lett.}\ }\textbf
  {\bibinfo {volume} {44}},\ \bibinfo {pages} {1502} (\bibinfo {year}
  {1980})}\BibitemShut {NoStop}%
\bibitem [{\citenamefont {Selke}(1988)}]{selke1988annni}%
  \BibitemOpen
  \bibfield  {author} {\bibinfo {author} {\bibfnamefont {W.}~\bibnamefont
  {Selke}},\ }\bibfield  {title} {\bibinfo {title} {{The ANNNI model --
  theoretical analysis and experimental application}},\ }\href
  {https://doi.org/10.1016/0370-1573(88)90140-8} {\bibfield  {journal}
  {\bibinfo  {journal} {Phys. Rep.}\ }\textbf {\bibinfo {volume} {170}},\
  \bibinfo {pages} {213} (\bibinfo {year} {1988})}\BibitemShut {NoStop}%
\bibitem [{\citenamefont {Wu}\ \emph {et~al.}(2019{\natexlab{c}})\citenamefont
  {Wu}, \citenamefont {Nikitin}, \citenamefont {Wang}, \citenamefont {Zhu},
  \citenamefont {Batista}, \citenamefont {Tsvelik}, \citenamefont {Samarakoon},
  \citenamefont {Tennant}, \citenamefont {Brando}, \citenamefont {Vasylechko},
  \citenamefont {Frontzek}, \citenamefont {Savici}, \citenamefont {Sala},
  \citenamefont {Ehlers}, \citenamefont {Christianson}, \citenamefont
  {Lumsden},\ and\ \citenamefont {Podlesnyak}}]{Wu2019Tomonaga}%
  \BibitemOpen
  \bibfield  {author} {\bibinfo {author} {\bibfnamefont {L.}~\bibnamefont
  {Wu}}, \bibinfo {author} {\bibfnamefont {S.}~\bibnamefont {Nikitin}},
  \bibinfo {author} {\bibfnamefont {Z.}~\bibnamefont {Wang}}, \bibinfo {author}
  {\bibfnamefont {W.}~\bibnamefont {Zhu}}, \bibinfo {author} {\bibfnamefont
  {C.}~\bibnamefont {Batista}}, \bibinfo {author} {\bibfnamefont
  {A.}~\bibnamefont {Tsvelik}}, \bibinfo {author} {\bibfnamefont
  {A.}~\bibnamefont {Samarakoon}}, \bibinfo {author} {\bibfnamefont
  {D.}~\bibnamefont {Tennant}}, \bibinfo {author} {\bibfnamefont
  {M.}~\bibnamefont {Brando}}, \bibinfo {author} {\bibfnamefont
  {L.}~\bibnamefont {Vasylechko}}, \bibinfo {author} {\bibfnamefont
  {M.}~\bibnamefont {Frontzek}}, \bibinfo {author} {\bibfnamefont
  {A.}~\bibnamefont {Savici}}, \bibinfo {author} {\bibfnamefont
  {G.}~\bibnamefont {Sala}}, \bibinfo {author} {\bibfnamefont {G.}~\bibnamefont
  {Ehlers}}, \bibinfo {author} {\bibfnamefont {A.}~\bibnamefont
  {Christianson}}, \bibinfo {author} {\bibfnamefont {M.}~\bibnamefont
  {Lumsden}},\ and\ \bibinfo {author} {\bibfnamefont {A.}~\bibnamefont
  {Podlesnyak}},\ }\bibfield  {title} {\bibinfo {title} {{Tomonaga-Luttinger
  liquid behavior and spinon confinement in YbAlO$_3$}},\ }\href
  {https://doi.org/10.1038/s41467-019-08485-7} {\bibfield  {journal} {\bibinfo
  {journal} {Nat. Commun.}\ }\textbf {\bibinfo {volume} {10}},\ \bibinfo
  {pages} {698} (\bibinfo {year} {2019}{\natexlab{c}})}\BibitemShut {NoStop}%
\bibitem [{\citenamefont {Nikitin}\ \emph {et~al.}(2021)\citenamefont
  {Nikitin}, \citenamefont {Nishimoto}, \citenamefont {Fan}, \citenamefont
  {Wu}, \citenamefont {Wu}, \citenamefont {Sukhanov}, \citenamefont {Brando},
  \citenamefont {Pavlovskii}, \citenamefont {Xu}, \citenamefont {Vasylechko},
  \citenamefont {Yu},\ and\ \citenamefont {Podlesnyak}}]{nikitin2021multiple}%
  \BibitemOpen
  \bibfield  {author} {\bibinfo {author} {\bibfnamefont {S.~E.}\ \bibnamefont
  {Nikitin}}, \bibinfo {author} {\bibfnamefont {S.}~\bibnamefont {Nishimoto}},
  \bibinfo {author} {\bibfnamefont {Y.}~\bibnamefont {Fan}}, \bibinfo {author}
  {\bibfnamefont {J.}~\bibnamefont {Wu}}, \bibinfo {author} {\bibfnamefont
  {L.}~\bibnamefont {Wu}}, \bibinfo {author} {\bibfnamefont {A.}~\bibnamefont
  {Sukhanov}}, \bibinfo {author} {\bibfnamefont {M.}~\bibnamefont {Brando}},
  \bibinfo {author} {\bibfnamefont {N.}~\bibnamefont {Pavlovskii}}, \bibinfo
  {author} {\bibfnamefont {J.}~\bibnamefont {Xu}}, \bibinfo {author}
  {\bibfnamefont {L.}~\bibnamefont {Vasylechko}}, \bibinfo {author}
  {\bibfnamefont {R.}~\bibnamefont {Yu}},\ and\ \bibinfo {author}
  {\bibfnamefont {A.}~\bibnamefont {Podlesnyak}},\ }\bibfield  {title}
  {\bibinfo {title} {{Multiple fermion scattering in the weakly coupled
  spin-chain compound YbAlO$_3$}},\ }\href
  {https://doi.org/10.1038/s41467-021-23585-z} {\bibfield  {journal} {\bibinfo
  {journal} {Nature Commun.}\ }\textbf {\bibinfo {volume} {12}},\ \bibinfo
  {pages} {1} (\bibinfo {year} {2021})}\BibitemShut {NoStop}%
\bibitem [{mti()}]{mtixtl}%
  \BibitemOpen
  \href@noop {} {}\bibinfo {howpublished} {https://www.mtixtl.com/}\BibitemShut
  {NoStop}%
\bibitem [{\citenamefont {Ehlers}\ \emph {et~al.}(2011)\citenamefont {Ehlers},
  \citenamefont {Podlesnyak}, \citenamefont {Niedziela}, \citenamefont
  {Iverson},\ and\ \citenamefont {Sokol}}]{CNCS1}%
  \BibitemOpen
  \bibfield  {author} {\bibinfo {author} {\bibfnamefont {G.}~\bibnamefont
  {Ehlers}}, \bibinfo {author} {\bibfnamefont {A.}~\bibnamefont {Podlesnyak}},
  \bibinfo {author} {\bibfnamefont {J.~L.}\ \bibnamefont {Niedziela}}, \bibinfo
  {author} {\bibfnamefont {E.~B.}\ \bibnamefont {Iverson}},\ and\ \bibinfo
  {author} {\bibfnamefont {P.~E.}\ \bibnamefont {Sokol}},\ }\bibfield  {title}
  {\bibinfo {title} {{The new cold neutron chopper spectrometer at the
  Spallation Neutron Source: design and performance}},\ }\href
  {https://doi.org/10.1063/1.3626935} {\bibfield  {journal} {\bibinfo
  {journal} {Rev. Sci. Instrum.}\ }\textbf {\bibinfo {volume} {82}},\ \bibinfo
  {pages} {085108} (\bibinfo {year} {2011})}\BibitemShut {NoStop}%
\bibitem [{\citenamefont {Ehlers}\ \emph {et~al.}(2016)\citenamefont {Ehlers},
  \citenamefont {Podlesnyak},\ and\ \citenamefont {Kolesnikov}}]{CNCS2}%
  \BibitemOpen
  \bibfield  {author} {\bibinfo {author} {\bibfnamefont {G.}~\bibnamefont
  {Ehlers}}, \bibinfo {author} {\bibfnamefont {A.}~\bibnamefont {Podlesnyak}},\
  and\ \bibinfo {author} {\bibfnamefont {A.~I.}\ \bibnamefont {Kolesnikov}},\
  }\bibfield  {title} {\bibinfo {title} {{The cold neutron chopper spectrometer
  at the Spallation Neutron Source - A review of the first 8 years of
  operation}},\ }\href {https://doi.org/10.1063/1.4962024} {\bibfield
  {journal} {\bibinfo  {journal} {Rev. Sci. Instrum.}\ }\textbf {\bibinfo
  {volume} {87}},\ \bibinfo {pages} {093902} (\bibinfo {year}
  {2016})}\BibitemShut {NoStop}%
\end{thebibliography}%
\end{document}